\documentclass[12pt]{article}

\usepackage[english]{babel}
\usepackage{epsfig}
\usepackage{amsmath}

\usepackage{graphicx}
\usepackage{geometry}
\usepackage[latin1]{inputenc}
\usepackage{amsmath,amssymb}

\newcommand{\pe}{\psi}
\def\d{\delta}
\def\ds{\displaystyle}
\def\e{{\epsilon}}
\def\eb{\bar{\eta}}
\def\enorm#1{\|#1\|_2}
\def\Fp{F^\prime}
\def\fishpack{{FISHPACK}}
\def\fortran{{FORTRAN}}
\def\gmres{{GMRES}}
\def\gmresm{{\rm GMRES($m$)}}
\def\Kc{{\cal K}}
\def\norm#1{\|#1\|}
\def\wb{{\bar w}}
\def\zb{{\bar z}}

\newcommand{\cav}{P}
\newcommand{\rr}{{\sf R}}
\newcommand{\eps}{\mathop{\varepsilon}\nolimits}
\newcommand{\dvr}{\rm div\;}
\newcommand{\rot}{\rm curl\;}
\newcommand{\grd}{\rm grad\,}
\newcommand{\sh}{\rm sh}
\newcommand{\ch}{\rm ch}
\newcommand{\summ}{\mbox{\Large$\Sigma$}}
\newcommand{\esssup}{\mathop{\rm ess\;sup}\nolimits}

\newtheorem{defin}{Definition}

\def\bfE{\mbox{\boldmath$E$}}
\def\bfG{\mbox{\boldmath$G$}}

\begin{document}



{\centerline{\Large{\bf{Eigenwaves in Waveguides with Dielectric
Inclusions:}}}}

{\centerline{\Large{\bf{Completeness}}}}
\bigskip
\bigskip
{\centerline{Y. Smirnov$^{\rm a}$ and Y. Shestopalov$^{\rm b}$}}
\bigskip
{\centerline{ $^{\rm a}$ Penza State University,
Penza, 440017 Russia (smirnov@penzadom.ru) }}

{\centerline{ $^{\rm b}$ Karlstad University, SE 65188 Karlstad,
Sweden (youri.shestopalov@kau.se) }}




\bigskip
\bigskip
\noindent {\bf{Abstract.}}\quad
We formulate the definition of eigenwaves and associated waves
in a nonhomogeneously filled waveguide
using the system of eigenvectors and associated vectors of a
pencil
and
prove its double completeness
with a finite defect or without a defect. Then we prove
%
the completeness of the system of transversal components of
eigenwaves and associated waves as well as
%
the `minimality' of this system
and show that this system is generally not a Schauder basis.
This work is a continuation of \cite{SS1}.
Therefore we omit the problem statements and all
necessary basic definitions
given in \cite{SS1}.
\bigskip

\noindent {\bf{Keywords:}}\quad eigenwave; waveguide; pencil;
spectrum; completeness; basis

\noindent {\bf{AMS Classification:}}\quad 45E99, 31B20,
83C50,74S10

\bigskip





\section{Introduction}
\label{section1}

Summarize the {\it new fundamental} results obtained in this study
for an arbitrary waveguide with nonhomogeneous filling and
arbitrary inclusions that belongs to the family described in
\cite{SS1}:

\noindent (i) the system of longitudinal components of normal
waves (eigenwaves and associated waves) is double complete
according to
Keldysh in the Sobolev spaces;

\noindent (ii) the system of transversal components of normal
waves (eigenwaves and associated waves) is complete in the space
of square-integrable functions.



We consider
properties of the system of eigenvectors and associated vectors of
a pencil $L\left( \gamma \right)$
for the problem on normal waves stated in
\cite{SS1}. We
establish, under certain conditions, double completeness of the
system
with a finite defect or without a defect
using perturbation techniques applied for a pencil of simple
structure and factorization of the pencil.

\section{Eigenvalue problem for the operator pencil}
\label{subsec:mylabel2w} Statement of the problem on normal waves
in a waveguide with nonhomogeneous filling is given in \cite{SS1},
Section 2, where all the necessary notations are introduced, as
well as the definition of the solution of the corresponding
boundary eigenvalue problem (6)--(9) (Definition 1) in terms of
its variational statement,
eq. (14) in Section 3 of \cite{SS1}. This problem
can be written in the operator form
for an
operator-valued pencil
\begin{eqnarray}
\label{eq19w} L\left( \gamma \right)f = 0,
\end{eqnarray}
\begin{eqnarray}
\nonumber
 L\left( \gamma \right): = \gamma ^4K + \gamma ^2\left( {A_1 - \left(
{\varepsilon _1 + \varepsilon _2 } \right)K} \right) + \left( {\varepsilon
_1 - \varepsilon _2 } \right)\gamma S + \varepsilon _1 \varepsilon _2 \left(
{K - A_2 } \right): H \to H,
\end{eqnarray}
where all the operators are bounded.
Eigenvalues and
eigenvectors of the pencil coincide with eigenvalues and
eigenfunctions of boundary eigenvalue problem (6)--(9),
\cite{SS1},
for $\gamma^2 \ne \varepsilon _1 $, $\gamma ^2 \ne \varepsilon _2
$. Thus the problem on normal waves is
reduced to an eigenvalue problem for pencil $L\left( \gamma
\right)$.

In \cite{SS1} we described main properties of operators of the
pencil and spectrum of the boundary eigenvalue problem under
study.

Denote by $\rho \left( L \right)$ the resolvent set of $L\left(
\gamma \right)$ (consisting of all complex values of $\gamma $ at
which there exists a bounded inverse operator $L^{ - 1}\left(
\gamma \right))$ and by $\sigma \left( L \right) = {\bf
C}\backslash \rho \left( L \right)$ the spectrum of $L\left(
\gamma \right)$.
The definitions of a holomorphic operator-function and
eigenvectors and associated vectors of a pencil
are given in \cite{SS1}.

\textbf{Definition 1}. \textit{ 
\label{def4} The system of eigenvectors and associated vectors of
operator-function $A\left( \gamma \right)$ is called complete with
power $n$ if any set of $n$ vectors $f_0 ,f_1 ,\dots ,f_{n - 1} $
can be represented as a limit with respect to the norm of the
linear combination of the elements of the system
\begin{equation}
\label{eq28w} f_{v,N} = \sum\limits_{k = 1}^N {\sum\limits_p
{a_{p,N}^{\left( k \right)} \varphi _p^{\left( {k,v} \right)} } }
, \quad v = 0,1,...,n - 1,
\end{equation}
where the coefficients do not depend on $v$,
\[
\varphi _p^{\left( {k,v} \right)} = \left. {\frac{d^v}{dt^v}} \right|_{t =
0} e^{\gamma _k t}\left( {\varphi _p^{\left( k \right)} + \varphi _{p -
1}^{\left( k \right)} \frac{t}{1!} + ... + \varphi _0^{\left( k \right)}
\frac{t^p}{p!}} \right),
\]
and $\gamma _k $ are eigenvalues of operator-function $A\left(
\gamma \right)$.}

For $n = 1$ the definition coincides with the standard definition
of the completeness of eigenvectors and associated vectors. If the
multiplicity of all eigenvectors is equal to 1 we have
\[
f_{v,N} = \sum\limits_{k = 1}^N {a_N^{\left( k \right)} \gamma _k^v \varphi
_0^{\left( k \right)} } .
\]

We will consider operator-functions $A\left( \gamma \right)$ that
have eigenvalues with finite algebraic multiplicity.

Let us study the pencil $L\left( \gamma \right)$. It
is more convenient to consider a regularized pencil
\begin{equation}
\label{eq29w}
 \tilde {L}\left( \gamma \right): = A_1^{ - 1 / 2} L A_1^{ - 1 / 2} = \gamma ^4\tilde {K} + \gamma ^2\left( {I - \left(
{\varepsilon _1 + \varepsilon _2 } \right)\tilde {K}} \right)
 + \left( {\varepsilon _1 - \varepsilon _2 } \right)\gamma \tilde {S} +
\varepsilon _1 \varepsilon _2 \left( {\tilde {K} - \tilde {A}_2 } \right),
\end{equation}
where $\tilde {K} = A_1^{ - 1 / 2} KA_1^{ - 1 / 2} $, $\tilde {S}
= A_1^{ - 1 / 2} SA_1^{ - 1 / 2} $, and $\tilde {A}_2 = A_1^{ - 1
/ 2} A_2 A_1^{ - 1 / 2} $.

It is easy to see that $\sigma \left( L \right) = \sigma \left(
\tilde {L} \right)$ and the following relations hold for
eigenvectors and associated vectors
\begin{equation}
\label{eq30w} \varphi _j \left( L \right) = A_1^{ - 1 / 2} \varphi
_j \left( \tilde {L} \right).
\end{equation}

Operators $\tilde {K}$, $\tilde {S}$ and $\tilde {A}_2 $ keep all
properties of operators $K$, $S$, and $A_2 $ given in \cite{SS1}.
\section{Completeness of the system of eigenvectors
and associated vectors of pencil $L\left( \gamma \right)$}
\label{section3}
We propose two approaches for the analysis of completeness of the
system of eigenvectors and associated vectors of pencil $L\left(
\gamma \right)$. Within the frames of the first approach, we
consider pencil $L\left( \gamma \right)$ as a perturbation of a
certain pencil of simple structure. We will analyze two cases. In
the first case, the original pencil is represented as a
perturbation of a Keldysh pencil by a holomorphic
operator-function; we will not impose any additional conditions
and prove only double completeness with a finite defect. In the
second case we prove double completeness of the system of
eigenvectors and associated vectors of pencil $L\left( \gamma
\right)$ under an additional condition that the parameter $\delta
= \left( {\varepsilon _2 - \varepsilon _1 } \right) / 2$ is
sufficiently small.

The second approach is based on factorization of pencil $L\left(
\gamma \right)$ with respect to a special contour on the complex
plane. We prove double completeness of the system of eigenvectors
and associated vectors of pencil $L\left( \gamma \right)$
corresponding to eigenvalues located outside  a certain contour.
However, we will impose conditions that govern parameters of the
pencil showing that these additional conditions are essential.

Note that $L\left( \gamma \right)$ does not belong to many
well-known families of pencils (Keldysh pencils, hyperbolic
pencils, and so on). In spite of this fact, we show that spectral
properties of the pencil can be efficiently studied.

Let us consider pencil $L\left( \gamma \right)$ in the domain
$D_\eta = \left\{ {\gamma :\left| \gamma \right| > \eta }
\right\}$ where $\eta $ is an arbitrary positive number such that
$\eta > \sqrt {\varepsilon _1 + \varepsilon _2 } $.
We have
\begin{equation}
\label{eq32w} F\left( \gamma \right): = \left( {\gamma ^2 - \left(
{\varepsilon _1 + \varepsilon _2 } \right)} \right)^{ - 1}\tilde
{L}\left( \gamma \right) = \gamma ^2\tilde {K} + I + \gamma ^{ -
1}T\left( \gamma \right),
\end{equation}
in the domain $D_\eta$,
where
\[
T\left( \gamma \right) = \gamma \left( {\gamma ^2 - \left( {\varepsilon _1 +
\varepsilon _2 } \right)} \right)^{ - 1} \left( {\left( {\varepsilon
_1 + \varepsilon _2 } \right)I + \left( {\varepsilon _1 - \varepsilon _2 }
\right)\gamma \tilde {S} + \varepsilon _1 \varepsilon _2 \left( {\tilde {K}
- \tilde {A}_2 } \right)} \right)
\]

Completeness
of the system of eigenvectors and associated vectors of pencil
$L\left( \gamma \right)$ corresponding to eigenvalues located in
$D_\eta $ is equivalent to completeness of the system of
eigenvectors and associated vectors of pencil $F\left( \gamma
\right)$ corresponding to eigenvalues located in $D_\eta $.
Indeed, the spectra of pencils in $D_\eta $ coincide and the
eigenvectors and associated vectors satisfy the relation
$
\varphi _j^{\left( k \right)} \left( L \right) = A_1^{ - 1 / 2} \varphi
_j^{\left( k \right)} \left( F \right),
$
which yields the equivalence of the problems of completeness of
systems $\left\{ {\varphi _j^{\left( k \right)} \left( L \right)}
\right\}$ and $\left\{ {\varphi _j^{\left( k \right)} \left( F
\right)} \right\}$.

\textbf{Theorem 1.} \textit{
\label{th5} The system of eigenvectors and associated vectors of
pencil $L\left( \gamma \right)$ corresponding to eigenvalues
located in domain $\left| \gamma \right| \ge \eta $ is double
complete with a finite defect in $H\times H:$
\[
\dim \mbox{coker}\;\overline {L\left( {\varphi _p^{\left( {k,0}
\right)} } \right)} < \infty , \quad \dim \mbox{coker}\;\overline
{L\left( {\varphi _p^{\left( {k,1} \right)} } \right)} < \infty ,
\]
where $\eta \ge 0$ is an arbitrary nonnegative number and
$\overline {L\left( {\varphi _p^{\left( {k,v} \right)} } \right)}
$ denotes the closure of linear combinations of vectors $\left\{
{\varphi _p^{\left( {k,v} \right)} } \right\}.$
}

{\bf{Proof.}}  It is sufficient to prove the theorem for pencil
$F\left( \gamma \right)$ under the condition $\eta > \sqrt
{\varepsilon _1 + \varepsilon _2 } $. Pencil $F\left( \gamma
\right)$ is considered as a perturbation of pencil $\gamma
^2\tilde {K} + I$ by the operator-function $T_1 \left( \gamma
\right) = \gamma ^{ - 1}T\left( \gamma \right)$, $T_1 \left(
\infty \right) = 0$, which is holomorphic in $D_\eta $. In this
case, $\tilde {K} > 0$ is a Hilbert-Schmidt operator;
consequently, all conditions of Theorem 1, \cite{147} are
fulfilled. This theorem implies double completeness of the system
of eigenvectors and associated vectors of pencil $F\left( \gamma
\right)$ (and $L\left( \gamma \right))$ with a finite defect in
$H\times H$; i.e., the closure of linear combinations of vectors $
\left( {\varphi _p^{\left( {k,0} \right)} ,\varphi _p^{\left(
{k,1} \right)} } \right)^{\rm T} \in H\times H, $ has a finite
defect in $H\times H$,$\gamma _k \in D_\eta $, where vectors
$\varphi _p^{\left( {k,v} \right)} $ are determined in Definition
1.

If we increase  $\eta $ then the dimension of the defect subspace
may also increase. On the other side, it is necessary to exclude
eigenvectors and associated vectors corresponding to eigenvalues
$\pm \sqrt {\varepsilon_i }$. Note however that the dimension of
the defect subspace is not known.

In applications, it is important to have a statement providing
completeness of the system of eigenvectors and associated vectors
of the pencil without defect. Below we formulate such a theorem
under the condition that the parameter $\delta = \left(
{\varepsilon _2 - \varepsilon _1 } \right) / 2$ is sufficiently
small.

Let us write $L\left( \gamma \right)$ in the form
\begin{eqnarray}
\label{eq39}
 L\left( \gamma \right) &=& \left( {\gamma ^2 - \frac{\varepsilon _1 +
\varepsilon _2 }{2}} \right)^2K - \left( {\frac{\varepsilon _1 -
\varepsilon _2 }{2}} \right)^2K + \left( {\gamma ^2 -
\frac{\varepsilon _1 + \varepsilon _2 }{2}} \right)A_1 + \\
\nonumber &+& \gamma \left( {\varepsilon _1 - \varepsilon _2 }
\right)S + \frac{\varepsilon _1 - \varepsilon _2 }{2}A'_1,
\end{eqnarray}
where $ \frac{\varepsilon _1 - \varepsilon _2 }{2}A'_1 =
\frac{\varepsilon _1 + \varepsilon _2 }{2}A_1 - \varepsilon _1
\varepsilon _2 A_2, $ and bounded self-adjoint operator $A'_1 $ is
described in \cite{SS1}. Expression for pencil $\tilde {L}\left(
\gamma \right)$ has the form
\begin{equation}
\label{eq40}
\tilde {L}\left( \gamma \right) = \left( {p^2 - \gamma ^2} \right)\left(
{\left( {p^2 - \gamma ^2} \right)\tilde {K} - I} \right) + \delta B\left(
\gamma \right),
\end{equation}
where
$
B\left( \gamma \right): = - 2\gamma \tilde {S} - \delta \tilde {K} - \tilde
{A}'_1 ;
\quad
p = \sqrt {{\varepsilon _1 + \varepsilon _2 }/{2}} .
$

The spectrum of pencil $L\left( \gamma \right)$ coincides with the
spectrum of pencil $\tilde {L}\left( \gamma \right)$. Eigenvectors
and associated vectors of the pencils satisfy formula
(\ref{eq30w}). Hence the completeness of system of eigenvectors
and associated vectors of pencil $L\left( \gamma \right)$ is
equivalent to that
of $\tilde {L}\left( \gamma \right)$.

\textbf{Theorem 2}. \textit{
\label{th6} Let $M > 1$ be an arbitrary number. Then there exists
a $\delta _\ast = \delta _\ast \left( {M;\Omega } \right)$ such
that for any $\varepsilon _j$ satisfying $ 1 \le \varepsilon _j
\le M$, the system of eigenvectors and associated vectors of
pencil $L\left( \gamma \right)$ corresponding to eigenvalues
$\gamma _n \ne \pm \sqrt {\varepsilon_i }$, $i = 1,\,2$, is double
complete in $H\times H$ under the condition $\left| \delta \right|
< \delta_\ast $.
}

{\bf{Proof.}} It is sufficient to prove that the theorem is valid
for pencil $\tilde {L}\left( \gamma \right)$. We consider $\tilde
{L}\left( \gamma \right)$ as a perturbation of the simple pencil
\begin{equation}
\label{eq41}
F_0 \left( \gamma \right) = \left( {p^2 - \gamma ^2} \right)\left( {\left(
{p^2 - \gamma ^2} \right)\tilde {K} - I} \right) = \left( {p^2 - \gamma ^2}
\right)\tilde {F}\left( \gamma \right)
\end{equation}
by operator-function $\delta B\left( \gamma \right)$, where
\begin{equation}
\label{eq42} \tilde {F}\left( \gamma \right) = \left( {p^2 -
\gamma ^2} \right)\tilde {K} - I.
\end{equation}

Spectrum $\sigma \left( \tilde {F} \right)$ is located on the real
and imaginary axes. The spectrum consists of eigenvalues of finite
algebraic multiplicity with an accumulation point at infinity (see
Fig. 3 in \cite{SS1}). Eigenvectors of pencil $\tilde {F}\left(
\gamma \right)$ form an orthonormal basis in $H$ (by the
Hilbert--Schmidt theorem \cite{109}). Since $\tilde {K} > 0$ the
eigenvalues $\tilde {\gamma }_n $ of pencil $\tilde {F}\left(
\gamma \right)$ satisfy the estimate $ \tilde {\gamma }_n^2 \le
p^2 - {\| \tilde {K} \|}^{-1}, \quad n = \pm 1,\pm 2,\dots , $
where $\tilde {\gamma }_{ - n} = - \tilde {\gamma }_n $ and the
numeration of eigenvalues decreases with respect to values $\tilde
{\gamma }_n^2 $. Under the conditions imposed on coefficients
$\varepsilon _j $ there exists an $M_0 < 1$ (which does not depend
on $\varepsilon _j )$ such that $p - \tilde {\gamma }_1 \ge 3M_0
$. Let $\left| \delta \right| \le M_0 $. Introduce the circles $
\Gamma _\pm = \left\{ {\gamma :\left| {\gamma \mp p} \right| =
r;\;r = p + M_0 - \sqrt {\varepsilon _{\min } } } \right\} $ and
consider pencil $\tilde {L}\left( \gamma \right)$ in the domain
$
D = \left\{ {\gamma :\left| {\gamma - p} \right| > r,\left| {\gamma + p}
\right| > r} \right\}.
$

The domain contains all eigenvalues of pencil $\tilde {F}\left(
\gamma \right)$. $\tilde {L}\left( \gamma \right)$ is a Fredholm
operator in $D$ \cite{SS1}. If $\gamma _0 \in \Gamma _0 $,
$\Gamma _0 = \Gamma _ + \cup \Gamma _ - $ then $\left| {\gamma _0
- \tilde {\gamma }_n } \right| \ge M_0 $, $\left| {\gamma _n \pm
\sqrt {\varepsilon _i } } \right| \ge M_0 $. Hence $F_0 \left(
\gamma \right)$ has a bounded inverse on $\Gamma _0 $ and $\left\|
{F_0^{ - 1} \left( \gamma \right)} \right\| \le C_0 $, where $C_0
$ does not depend on $\varepsilon _j $, $\gamma $. Moreover,
$\left\| {B\left( \gamma \right)} \right\| \le B_0 $ on $\Gamma _0
$. If $\left| \delta \right| < B_0^{ - 1} C_0^{ - 1} $ then
$\tilde {L}\left( \gamma \right)$ also has a bounded inverse on
$\Gamma _0 $ and $\left\| {\tilde {L}_{ - 1} \left( \gamma
\right)} \right\| \le C_1 $ uniformly with respect to $\gamma \in
\Gamma _0 $, $1 \le \varepsilon _j \le M$.

According to \cite{147} in order to prove double completeness of
the system of eigenvectors and associated vectors of pencil
$\tilde {L}\left( \gamma \right)$ corresponding to eigenvalues
located in $D$ it is sufficient to show that if the
vector-function $f\left( \gamma \right) = \tilde {L}^{ - 1}\left(
\gamma \right)\left( {f_0 + \gamma f_1 } \right)$ is holomorphic
in $D$ then $f_0 = f_1 = 0$ for any $f_0 ,\,f_1 \in H$.

Let $f\left( \gamma \right)$ be holomorphic in $D$. It follows
from \cite{SS1} that the operator-function
\[
F^{ - 1}\left( \gamma \right)\left( {\gamma ^{ - 1}f_0 + f_1 } \right) =
{\gamma }^{-1}({\gamma ^2 - \varepsilon _1 - \varepsilon _2 })\tilde {L}^{ -
1}\left( \gamma \right)\left( {f_0 + \gamma f_1 } \right)
\]
is bounded at infinity and the following expansion holds
\[
\tilde {L}^{ - 1}\left( \gamma \right)\left( {f_0 + \gamma f_1 } \right) =
\sum\limits_{k = 1}^\infty {g_k \gamma ^{ - k}} ,
\quad
\left| \gamma \right| > R.
\]
From the equality
\[
f_0 + \gamma f_1 = \tilde {L}\left( \gamma \right)\tilde {L}^{ - 1}\left(
\gamma \right)\left( {f_0 + \gamma f_1 } \right) = \left( {\gamma ^4\tilde
{K} + ...} \right)\sum\limits_{k = 1}^\infty {g_k \gamma ^{ - k}}
\]
and the property $\tilde {K} > 0$ we have $g_1 = g_2 = 0$;
consequently $f\left( \gamma \right)$ has a zero at infinity of
the order not less than 3. Let us integrate the equalities
\[
\gamma F_0^{ - 1} \left( \gamma \right)\left( {f_0 + \gamma f_1 } \right) -
\gamma f\left( \gamma \right) = \gamma F_0^{ - 1} \left( \gamma
\right)\delta B\left( \gamma \right)f\left( \gamma \right),
\]
\[
F_0^{ - 1} \left( \gamma \right)\left( {f_0 + \gamma f_1 } \right) - f\left(
\gamma \right) = F_0^{ - 1} \left( \gamma \right)\delta B\left( \gamma
\right)f\left( \gamma \right)
\]
with respect to $\Gamma _0 $. Taking into account the properties
of $f\left( \gamma \right)$ at infinity and continuous
invertibility of $\tilde {L}\left( \gamma \right)$ and $F_0 \left(
\gamma \right)$ on $\Gamma _0 $ we obtain
\[
f_0 = \frac{\delta }{2\pi i}\int\limits_{\Gamma _0 } {\gamma F_0^{ - 1}
\left( \gamma \right)B\left( \gamma \right)f\left( \gamma \right)d\gamma }
, \quad
f_1 = \frac{\delta }{2\pi i}\int\limits_{\Gamma _0 } {F_0^{ - 1} \left(
\gamma \right)B\left( \gamma \right)f\left( \gamma \right)d\gamma } .
\]
Hence there exists a $c > 0$ (which does not depend on
$\varepsilon _j $ and $\gamma $) such that
$
\left\| {f_0 } \right\| \le \left| \delta \right|C\left( {\left\| {f_0 }
\right\| + \left\| {f_1 } \right\|} \right),
$
$
\left\| {f_1 } \right\| \le \left| \delta \right|C\left( {\left\| {f_0 }
\right\| + \left\| {f_1 } \right\|} \right).
$
If $\left| \delta \right| < \left( {2C} \right)^{ - 1}$ then
$\left\| {f_0 } \right\| = \left\| {f_1 } \right\| = 0$ and $f_0 =
f_1 = 0$.

In order to complete the proof we can choose
$
\delta _\ast < \min \left( {\left( {2C} \right)^{ - 1},M_0 ,B_0^{ - 1} C_0^{
- 1} } \right).
$

Now we  prove completeness of the system of eigenvectors and
associated vectors of pencil $L\left( \gamma \right)$ using the
factorization method.

\textbf{Theorem 3.} \textit{
\label{th7} The system of eigenvectors and associated vectors of
pencil $L\left( \gamma \right)$ corresponding to eigenvalues
$\gamma_n \ne \pm \sqrt {\varepsilon_i }$, $i = 1,2$, is double
complete in $H\times H$ under the conditions
\begin{equation}
\label{eq43}
\varepsilon _{\max } < 9\varepsilon _{\min }
\end{equation}
and
\begin{equation}
\label{eq44}
\int\limits_\Omega {\left( {\varepsilon \left| \Pi \right|^2 + \left| \Psi
\right|^2} \right)dx} \le \frac{1}{2}\int\limits_\Omega {\left( {\left|
{\nabla \Pi } \right|^2 + \frac{1}{\varepsilon }\left| {\nabla \Psi }
\right|^2} \right)dx} ,\quad \forall \Pi \in H_0^1 \left( \Omega \right),
\quad
\Psi \in \mathord{\buildrel{\lower3pt\hbox{$\scriptscriptstyle\frown$}}\over
{H}} ^1\left( \Omega \right).
\end{equation}
}

{\bf{Proof.}} In order to factorize pencil $L\left( \gamma
\right)$ let us determine domains on the plane of variable $\gamma
$, where the equation
\begin{equation}
\label{eq45}
\left( {L\left( \gamma \right)f,f} \right) = 0
\end{equation}
has a definite number of roots with respect to $\gamma $ for any
$f \ne 0$. First, let us prove two estimates. Using Green's
formula and Schwartz inequality we obtain:
\[
\left| {\int\limits_\Gamma {\left( {\frac{\partial \Pi }{\partial \tau }\bar
{\Psi } - \frac{\partial \Psi }{\partial \tau }\bar {\Pi }} \right)d\tau } }
\right| = \left| {\int\limits_{\Omega _j } {\left( {\frac{\partial \Pi
}{\partial x_2 }\frac{\partial \bar {\Psi }}{\partial x_1 } - \frac{\partial
\Pi }{\partial x_1 }\frac{\partial \bar {\Psi }}{\partial x_2 } +
\frac{\partial \Psi }{\partial x_1 }\frac{\partial \bar {\Pi }}{\partial x_2
} - \frac{\partial \Psi }{\partial x_2 }\frac{\partial \bar {\Pi }}{\partial
x_1 }} \right)dx} } \right| =
\]
\[
 = 2\left| {\int\limits_{\Omega _j } {Re\left( {\frac{\partial \Pi
}{\partial x_2 }\frac{\partial \bar {\Psi }}{\partial x_1 } - \frac{\partial
\Pi }{\partial x_1 }\frac{\partial \bar {\Psi }}{\partial x_2 }} \right)dx}
} \right| \le 2\left| {\int\limits_{\Omega _j } {\left( {\frac{\partial \Pi
}{\partial x_2 }\frac{\partial \bar {\Psi }}{\partial x_1 } - \frac{\partial
\Pi }{\partial x_1 }\frac{\partial \bar {\Psi }}{\partial x_2 }} \right)dx}
} \right| \le
\]
\[
 \le 2\left( {\int\limits_{\Omega _j } {\left| {\nabla \Pi } \right|^2dx} }
\right)^{1 / 2}\left( {\int\limits_{\Omega _j } {\left| {\nabla \Psi }
\right|^2dx} } \right)^{1 / 2};
j = 1,2.
\]
Thus
\begin{equation}
\label{eq46}
\left| {\int\limits_\Gamma {\left( {\frac{\partial \Pi }{\partial \tau }\bar
{\Psi } - \frac{\partial \Psi }{\partial \tau }\bar {\Pi }} \right)d\tau } }
\right|^2 \le 4\int\limits_{\Omega _j } {\left| {\nabla \Pi } \right|^2dx}
\int\limits_{\Omega _j } {\left| {\nabla \Psi } \right|^2dx} .
\end{equation}

Consider the equation (w.r.t. $\sqrt {\varepsilon _j } $)
$
\varepsilon _j P + Q\pm \sqrt {\varepsilon _j } R = 0,
$
where
\[
P = \int\limits_{\Omega _j } {\left| {\nabla \Pi } \right|^2dx} \;\left( {
\ge 0} \right),
\quad
Q = \int\limits_{\Omega _j } {\left| {\nabla \Psi } \right|^2dx} \,\left( {
\ge 0} \right),
\quad
R = \int\limits_\Gamma {\left( {\frac{\partial \Pi }{\partial \tau }\bar
{\Psi } - \frac{\partial \Psi }{\partial \tau }\bar {\Pi }} \right)d\tau }
.
\]
Using estimate (\ref{eq46}) we have $ R^2 - 4PQ \le 0$ and for any
$\sqrt {\varepsilon _j } $ inequality
$
\varepsilon _j P + Q\pm \sqrt {\varepsilon _j } R \ge 0,
$
which is equivalent to the inequality
\begin{equation}
\label{eq47}
\int\limits_{\Omega _j } {\left( {\varepsilon _j \left| {\nabla \Pi }
\right|^2 + \left| {\nabla \Psi } \right|^2} \right)dx} \pm \sqrt
{\varepsilon _j } \int\limits_\Gamma {\left( {\frac{\partial \Pi }{\partial
\tau }\bar {\Psi } - \frac{\partial \Psi }{\partial \tau }\bar {\Pi }}
\right)d\tau } \ge 0,
\quad
j = 1,2.
\end{equation}

Assume that $\varepsilon _2 \ge \varepsilon _1 $. Denote
\[
s: = \int\limits_\Gamma {\left( {\frac{\partial \Pi }{\partial \tau }\bar
{\Psi } - \frac{\partial \Psi }{\partial \tau }\bar {\Pi }} \right)d\tau } ,
\quad
k: = \int\limits_\Omega {\left( {\varepsilon \left| \Pi \right|^2 + \left|
\Psi \right|^2} \right)dx} ,
\]
\[
a^{\left( j \right)}: = \int\limits_{\Omega _j } {\left( {\varepsilon _j
\left| {\nabla \Pi } \right|^2 + \left| {\nabla \Psi } \right|^2} \right)dx}
,\quad
a_2 : = \int\limits_\Omega {\left( {\left| {\nabla \Pi } \right|^2 +
\frac{1}{\varepsilon }\left| {\nabla \Psi } \right|^2} \right)dx} .
\]
Estimates (\ref{eq47}) have the form
\begin{equation}
\label{eq48}
a^{\left( 1 \right)}\pm \sqrt {\varepsilon _1 } s \ge 0,
\quad
a^{\left( 2 \right)}\pm \sqrt {\varepsilon _2 } s \ge 0.
\end{equation}
According to eq. (12) in \cite{SS1}
equation (\ref{eq45}) can be
represented as
\begin{equation}
\label{eq49}
f\left( \gamma \right): = \frac{a^{\left( 1 \right)} + \gamma s}{\varepsilon
_1 - \gamma ^2} + \frac{a^{\left( 2 \right)} - \gamma s}{\varepsilon _2 -
\gamma ^2} = k,
\quad
\gamma ^2 \ne \varepsilon _j ,
\end{equation}
where $a^{\left( 1 \right)} \ge 0$, $a^{\left( 2 \right)} \ge 0$, and
$k > 0$.

If $\gamma \in \left( { - \sqrt {\varepsilon_1 } ,\sqrt
{\varepsilon_1 } } \right)$ then $( {a^{\left( 1 \right)} \ne 0}
)$
\begin{eqnarray*}
 \frac{a^{\left( 1 \right)} + \gamma s}{\varepsilon _1 - \gamma ^2}
 &=&
\frac{a^{\left( 1 \right)} + \sqrt {\varepsilon _1 } s}{2\sqrt {\varepsilon
_1 } \left( {\sqrt {\varepsilon _1 } - \gamma } \right)} + \frac{a^{\left( 1
\right)} - \sqrt {\varepsilon _1 } s}{2\sqrt {\varepsilon _1 } \left( {\sqrt
{\varepsilon _1 } + \gamma } \right)}
 \ge \\ &\ge & \min \left( {\frac{a^{\left( 1 \right)}}{\sqrt {\varepsilon _1 } \left(
{\sqrt {\varepsilon _1 } - \gamma } \right)},\;\frac{a^{\left( 1
\right)}}{\sqrt {\varepsilon _1 } \left( {\sqrt {\varepsilon _1 }
+ \gamma } \right)}} \right) > \frac{a^{\left( 1
\right)}}{2\varepsilon _1 }.
\end{eqnarray*}
If $\gamma \in \left( { - \sqrt {\varepsilon _2 }
,\sqrt {\varepsilon _2 } } \right)$ then $( {a^{\left( 2 \right)}
\ne 0} )$
\begin{eqnarray*}
 \frac{a^{\left( 2 \right)} - \gamma s}{\varepsilon _2 - \gamma ^2}
 &=& \frac{a^{\left( 2 \right)} - \sqrt {\varepsilon _2 } s}{2\sqrt
{\varepsilon _2 } \left( {\sqrt {\varepsilon _2 } - \gamma }
\right)} + \frac{a^{\left( 2 \right)} + \sqrt {\varepsilon _2 }
s}{2\sqrt {\varepsilon _2 } \left( {\sqrt {\varepsilon _2 } +
\gamma } \right)}
 \ge \\&\ge &\min \left( {\frac{a^{\left( 2 \right)}}{\sqrt {\varepsilon _2 } \left(
{\sqrt {\varepsilon _2 } + \gamma } \right)},\;\frac{a^{\left( 2
\right)}}{\sqrt {\varepsilon _2 } \left( {\sqrt {\varepsilon _2 }
- \gamma } \right)}} \right) > \frac{a^{\left( 2
\right)}}{2\varepsilon _2 }. \end{eqnarray*}

Thus for $\gamma \in \left( { - \sqrt {\varepsilon _1 } ,\sqrt {\varepsilon
_1 } } \right)$ we have the estimate
\[
f\left( \gamma \right) > \frac{1}{2}\left( {\frac{a^{\left( 1
\right)}}{\varepsilon _1 } + \frac{a^{\left( 2 \right)}}{\varepsilon _2 }}
\right) = \frac{a_2 }{2}.
\]
From this estimate and the conditions of Theorem 3
it follows that equation (\ref{eq49}) (and (\ref{eq45})) has no
real roots for $\gamma \in \left( { - \sqrt {\varepsilon _1 }
,\sqrt {\varepsilon _1 } } \right)$. From the representation
\[
 f\left( \gamma \right) = \frac{a^{\left( 1 \right)} + \sqrt {\varepsilon _1
} s}{2\sqrt {\varepsilon _1 } \left( {\sqrt {\varepsilon _1 } - \gamma }
\right)} + \frac{a^{\left( 1 \right)} - \sqrt {\varepsilon _1 } s}{2\sqrt
{\varepsilon _1 } \left( {\sqrt {\varepsilon _1 } + \gamma } \right)}
 + \frac{a^{\left( 2 \right)} - \sqrt {\varepsilon _2 } s}{2\sqrt
{\varepsilon _2 } \left( {\sqrt {\varepsilon _2 } - \gamma } \right)} +
\frac{a^{\left( 2 \right)} + \sqrt {\varepsilon _2 } s}{2\sqrt {\varepsilon
_2 } \left( {\sqrt {\varepsilon _2 } + \gamma } \right)},
\]
$\gamma^2 \ne \varepsilon_j$, we obtain the following property. If
the sign in inequalities (\ref{eq48}) is $>$, then equation
(\ref{eq49}) has at least one root on each interval $\left( { -
\sqrt {\varepsilon _2 } , - \sqrt {\varepsilon _1 } } \right)$ and
$\left( {\sqrt {\varepsilon _1 } ,\sqrt {\varepsilon _2 } }
\right)$, and there are no roots of the equation for $\gamma \in
\left( { - \infty , - \sqrt {\varepsilon _2 } } \right) \cup
\left( {\sqrt {\varepsilon _2 } , + \infty } \right)$. Here we
have taken into account the properties
\[
\mathop {\lim }\limits_{\gamma \to \sqrt {\varepsilon _j } \mp 0}
f\left( \gamma \right) = \mathop {\lim }\limits_{\gamma \to -
\sqrt {\varepsilon _j } \pm 0} f\left( \gamma \right) = \pm \infty
,(j = 1,2)\quad {\mbox {and}} \quad f\left( \gamma \right) < 0,
\quad \left| \gamma \right| > \sqrt {\varepsilon _2 } .
\]

Let $\gamma _1 $ and $\gamma _2 $ be the roots of equation
(\ref{eq45}). We can calculate another two roots of (\ref{eq45})
by the Viete formula:
\begin{equation}
\label{eq50}
 \gamma _{3,4} = - \frac{\left( {\gamma _1 + \gamma _2 } \right)}{2}\pm
i\frac{\sqrt {4\varepsilon _1 \varepsilon _2 \theta / \left| {\gamma _1
\gamma _2 } \right| - \left( {\gamma _1 + \gamma _2 } \right)^2} }{2},
\end{equation}
where ${\displaystyle \theta : = {a_2 }/{k} - 1(1 \le \theta <
+ \infty )}$.

Using the inequalities ${\displaystyle {1}/{\left| {\gamma _1
\gamma _2 } \right|} \ge {1}/{\varepsilon _2 }}$ and
${\displaystyle \left| {\gamma _1 + \gamma _2 } \right| \le \sqrt
{\varepsilon _2 } - \sqrt {\varepsilon _1 } }$ it is easy to
verify that there exists $\tilde {\delta } > 0$ (which depends on
$\varepsilon _j )$ such that for $\varepsilon_2 < 9\varepsilon_1 $
we have
\begin{equation}
\label{eq51}
\left| {\gamma _{3,4} - p} \right| > \tilde {r};
\tilde {r} = \frac{\sqrt {\varepsilon _2 } - \sqrt {\varepsilon _1 } }{2} +
\tilde {\delta },
p = \frac{\sqrt {\varepsilon _1 } + \sqrt {\varepsilon _2 } }{2}.
\end{equation}
Thus in the domains
$
\left\{ {\gamma :\left| {p - \gamma } \right| < \tilde {r}} \right\},
\quad
\left\{ {\gamma :\left| {p + \gamma } \right| < \tilde {r}} \right\},
$
equation (\ref{eq45}) has only one real root on each interval $\left( { - \sqrt
{\varepsilon _2 } , - \sqrt {\varepsilon _1 } } \right)$ and $\left( {\sqrt
{\varepsilon _1 } ,\sqrt {\varepsilon _2 } } \right)$, and the equation has
two roots in the domain
$
\left\{ {\gamma :\left| {p\pm \gamma } \right| > \tilde {r}} \right\}.
$

Taking into account the continuity of roots of equation
(\ref{eq50}) with respect to the coefficients we find that $\gamma
_1 \in \left[ { - \sqrt {\varepsilon _2 } , - \sqrt {\varepsilon
_1 } } \right]$, $\gamma _2 \in \left[ {\sqrt {\varepsilon _1 }
,\sqrt {\varepsilon _2 } } \right]$, and $\gamma _{3,4} \in
\left\{ {\gamma :\left| {p\pm \gamma } \right| \ge \tilde {r}}
\right\}$ under conditions (\ref{eq48}).

Choose $\delta _0 = {\tilde {\delta }}/{2}$ and denote $r = ({\sqrt
{\varepsilon _2 } - \sqrt {\varepsilon _1 } })/{2} + \delta _0 $,
$
\sigma _ + = \left\{ {\gamma :\left| {p - \gamma } \right| < r} \right\}
$,
$
\sigma _ - = \left\{ {\gamma :\left| {p + \gamma } \right| < r} \right\}
$,
$
\sigma _1 = C\backslash \left( {\overline {\sigma _ + } \cup \overline
{\sigma _ - } } \right)
$,
$
\sigma _2 = \sigma _ + \cup \sigma _ - .
$

We proved that domains $\sigma_+$ and $\sigma_-$ contain only one
root of equation (\ref{eq45}) and domain $\sigma_1 $ contains two
roots of (\ref{eq45}). Spectrum of pencil $L\left( \gamma \right)$
is divided into three domains $\sigma _ + $, $\sigma _- $, and
$\sigma _1 $. We have also $ \sigma \left( L \right) \cap \sigma _
- \subset \left[ { - \sqrt {\varepsilon _2 } , - \sqrt
{\varepsilon _1 } } \right]$, $\sigma \left( L \right) \cap \sigma
_ + \subset \left[ {\sqrt {\varepsilon _1 } ,\sqrt {\varepsilon _2
} } \right]. $ Let $ \Gamma _\pm = \left\{ {\gamma :\left| {\gamma
\mp p} \right| = r} \right\}$, $\Gamma_1 = \Gamma_+ \cup \Gamma_-
$. Equation (\ref{eq45}) has not roots on $\Gamma _1 $ and
\begin{equation}
\label{eq52}
\mathop {\inf }\limits_{\left\| f \right\| = 1,\,\gamma \in \Gamma _1 }
\left| {\left( {L\left( \gamma \right)f,f} \right)} \right| > 0.
\end{equation}

Estimate (\ref{eq52}) and properties of the spectrum of pencil
$L\left( \gamma \right)$ remain valid for pencil $\tilde {L}\left(
\gamma \right)$. In this case the pencil $\tilde {L}\left( \gamma
\right)$ admits factorization with respect to contour $\Gamma _1
$:
\begin{equation}
\label{eq53}
\tilde {L}\left( \gamma \right) = L_1 \left( \gamma \right)L_2 \left( \gamma
\right),
\end{equation}
where
\[
L_1 \left( \gamma \right) = \gamma ^2\tilde {K} + \gamma \tilde {K}B_1 + I +
\tilde {K}\left( {\left( {\varepsilon _1 + \varepsilon _2 } \right)I + B_2 -
B_1^2 } \right),
\quad
L_2 \left( \gamma \right) = \gamma ^2I + \gamma B_1 + B_2 ,
\]
$B_1 $ and $B_2 $ are bounded operators and
$
\sigma \left( {L_1 } \right) \subset \sigma _1 ,
$,
$
\sigma \left( {L_2 } \right) \subset \sigma _2 .$

This assertion follows from \cite{128} (note that in order to use
Theorems 1 and 2 of \cite{128} it is necessary to introduce the
new variable $t = \left( {p - \gamma } \right)^{ - 1})$.


Pencil $L_2 \left( \gamma \right)$ has a bounded inverse on
$\overline {\sigma _1 } $. The system of eigenvectors and
associated vectors of pencil $L_1 \left( \gamma \right)$ is double
complete in $H\times H$ by the Keldysh theorem \cite{107} and,
consequently, the system of eigenvectors and associated vectors of
pencil $\tilde {L}\left( \gamma \right)$ corresponding to the
eigenvalues $\gamma _n \in \sigma _1 $ is also double complete in
$H\times H$. The completeness of the system of eigenvectors and
associated vectors of $L\left( \gamma \right)$ is equivalent to
the completeness of the system of eigenvectors and associated
vectors of $\tilde {L}\left( \gamma \right)$.

Note that the equivalence of reduction of the boundary value
problem on normal waves to an eigenvalue problem of the pencil at
the points $\gamma _j = \pm \sqrt {\varepsilon _i } $ is not
valid. In Theorems 2 and 3
we formulate the
conditions which separate degeneration points $\pm \sqrt
{\varepsilon _i } $ of pencil $L\left( \gamma \right)$. We will
not consider the system of eigenvectors and associated vectors of
pencil $L\left( \gamma \right)$ corresponding to eigenvalues $\pm
\sqrt {\varepsilon _i } $. Theorem 1
shows that the
system of eigenvectors and associated vectors of pencil $L\left(
\gamma \right)$ corresponding to the eigenvalues satisfying the
condition $\left| \gamma \right| > \eta $ for arbitrary $\eta > 0$
is in fact sufficiently broad. We can add only a finite set of
elements in order to obtain a double complete system in $H\times
H$.

Below we will show that double completeness of the system of
eigenvectors and associated vectors of pencil $L\left( \gamma
\right)$ is important for the analysis of the boundary value
problem on normal waves. Other statements concerning completeness
of the system of eigenvectors and associated vectors of pencil
$L\left( \gamma \right)$ can be found in \cite{94, 165}.
\section{Properties of the system of eigenwaves and
associated waves of a waveguide} \label{section4} The section is
devoted to investigation of the properties of the system of
eigenwaves and associated waves of the waveguide considered in
Section
\ref{section3}:
completeness, basis property, and biorthogonality.
These properties are of crucial importance when excitation of
waveguides is considered
which is reduced to
nonhomogeneous boundary value problems for the Maxwell and
Helmholtz equations. Note for example that if the completeness and
basis property of the system of eigenwaves and associated waves
are not established then the expansions of solutions to the
excitation problems \cite{63} are not correct.

We will consider only the case $\varepsilon _1 \ne \varepsilon _2
$ which ends up with a vector problem. For $\varepsilon _1 =
\varepsilon _2 $ the problem on normal waves is reduced to two
well-known scalar problems.

The results of this section are based on the methods and
techniques developed mainly in \cite{162, 165, 174, 163}.
\subsection{Eigenwaves and associated waves}
\label{subsec:mylabel1} Below we will use the notations of Section
2, \cite{SS1}.
Let $f_0 ,\;f_1 ,\;...,\;f_m \in H$ be the chain of
eigenvectors and associated vectors of pencil $L\left( \gamma
\right)$ corresponding to the eigenvalues $\gamma $ ($\gamma ^2
\ne \varepsilon _i $, $i = 1,2)$. Using the vectors $f_p = \left(
{\Pi _p ,\Psi _p } \right)^{\rm T}$ we define a system of functions on
$\Omega $:
\[
E_1^{\left( p \right)} = \frac{i\gamma }{\tilde {k}^2}\left(
{\frac{\partial \Pi _p }{\partial x_1 } - iE_1^{\left( {p - 1}
\right)} } \right) - \frac{i}{\tilde {k}^2}\left( {\frac{\partial
\Psi _p }{\partial x_2 } - iH_2^{\left( {p - 1} \right)} }
\right),
\]
\begin{equation}
\label{eq1} E_2^{\left( p \right)} = \frac{i\gamma }{\tilde
{k}^2}\left( {\frac{\partial \Pi _p }{\partial x_2 } -
iE_2^{\left( {p - 1} \right)} } \right) + \frac{i}{\tilde
{k}^2}\left( {\frac{\partial \Psi _p }{\partial x_1 } -
iH_1^{\left( {p - 1} \right)} } \right),
\end{equation}
\[
H_1^{\left( p \right)} = \frac{i\varepsilon }{\tilde {k}^2}\left(
{\frac{\partial \Pi _p }{\partial x_2 } - iE_2^{\left( {p - 1}
\right)} } \right) + \frac{i\gamma }{\tilde {k}^2}\left(
{\frac{\partial \Psi _p }{\partial x_1 } - iH_1^{\left( {p - 1}
\right)} } \right),
\]
\[
 H_2^{\left( p \right)} = - \frac{i\varepsilon }{\tilde
{k}^2}\left( {\frac{\partial \Pi _p }{\partial x_1 } -
iE_1^{\left( {p - 1} \right)} } \right) + \frac{i\gamma }{\tilde
{k}^2}\left( {\frac{\partial \Psi _p }{\partial x_2 } -
iH_2^{\left( {p - 1} \right)} } \right),
\]
\[
E_3^{\left( p \right)} = \Pi _p , \, H_3^{\left( p \right)} = \Psi
_p ; \quad E^{\left( p \right)} \equiv H^{\left( p \right)} \equiv
0\,\,{\mbox{for}}\,\, p < 0; \quad \varepsilon = \varepsilon _j
\,\, {\mbox{in}}\,\,\Omega _j,\,\, j = 1,2.
\]

\textbf{Definition 2}. \textit{
\label{def01}
The vector
\[
\tilde {W}^{\left( p \right)} = \tilde {V}^{\left( p \right)}\exp
\left( {i\gamma x_3 } \right), \quad \tilde {V}^{\left( p \right)}
= \left( {E_1^{\left( p \right)} ,E_2^{\left( p \right)}
,E_3^{\left( p \right)} ,H_1^{\left( p \right)} ,H_2^{\left( p
\right)} ,H_3^{\left( p \right)} } \right)^{\rm T},
\]
is called eigenwave for $p = 0$ or associated wave for $p \ge 1$
corresponding to eigenvalue $\gamma$.  }

Vector $\tilde {V}^{\left( p \right)}$ will be considered as an
element of the space
\[
\tilde {H} = L_2 \left( \Omega \right)\times L_2 \left( \Omega
\right)\times H_0^1 \left( \Omega \right)\times L_2 \left( \Omega
\right)\times L_2 \left( \Omega \right)\times
\mathord{\buildrel{\lower3pt\hbox{$\scriptscriptstyle\frown$}}\over
{H}}^1 \left( \Omega \right)
\]
with the standard inner product and norm defined for the product
of spaces.

From the results of Section 2
of \cite{SS1} it follows that $x_3$-components of vector $\tilde
{V}^{\left( 0 \right)}$ are eigenfunctions of the problem on
normal waves of a waveguide. Consequently, the above definition of
eigenwaves coincides with the standard definition. However, the
definition of an associated wave ($p \ge 1)$ is not a standard
one. We define an associated wave using associated vectors of
pencil $L\left( \gamma \right)$; the latter, in the general case,
is not connected directly with a boundary value problem for
Maxwell equations (see eq. (1) in \cite{SS1})
The standard
way of introducing associated waves is as follows. System of
Maxwell equations
can be considered as an eigenvalue problem
for the linear pencil
\[
M\left( \gamma \right) = M_1 + \gamma M_2 ,
\]
and eigenwaves and associated waves can be considered as a
solution of the boundary value problems
\[
\left( {M_1 + \gamma M_2 } \right)\dot {V}^{\left( 0 \right)} = 0,
\quad \left( {M_1 + \gamma M_2 } \right)\dot {V}^{\left( p
\right)} + M_2 \dot {V}^{\left( {p - 1} \right)} = 0, \quad p \ge
1,
\]
with the corresponding boundary and transmission conditions. In
terms of the Maxwell equations considered componentwise these problems
take the form
\[
\frac{\partial \dot {H}_3^{\left( p \right)} }{\partial x_2 } -
i\gamma \dot {H}_2^{\left( p \right)} - i\varepsilon \dot
{E}_1^{\left( p \right)} = i\dot {H}_2^{\left( {p - 1} \right)} , \quad
i\gamma \dot {H}_1^{\left( p \right)} - \frac{\partial \dot
{H}_3^{\left( p \right)} }{\partial x_1 } - i\varepsilon \dot
{E}_2^{\left( p \right)} = - i\dot {H}_1^{\left( {p - 1} \right)}
,
\]
\begin{equation}
\label{eq2}
\frac{\partial \dot {H}_2^{\left( p \right)} }{\partial x_1 } -
\frac{\partial \dot {H}_1^{\left( p \right)} }{\partial x_2 } -
i\varepsilon \dot {E}_3^{\left( p \right)} = 0, \quad
\frac{\partial \dot {E}_3^{\left( p \right)}
}{\partial x_2 } - i\gamma \dot {E}_2^{\left( p \right)} + i\dot
{H}_1^{\left( p \right)} = i\dot {E}_2^{\left( {p - 1} \right)} ,
\end{equation}
\[
i\gamma \dot {E}_1^{\left( p \right)} - \frac{\partial \dot
{E}_3^{\left( p \right)} }{\partial x_1 } + i\dot {H}_2^{\left( p
\right)} = - i\dot {E}_1^{\left( {p - 1} \right)} , \quad
\frac{\partial \dot {E}_2^{\left( p \right)} }{\partial x_1 } -
\frac{\partial \dot {E}_1^{\left( p \right)} }{\partial x_2 } +
i\dot {H}_3^{\left( p \right)} = 0;
\]
\begin{equation}
\label{eq3} \left. {\dot {E}_\tau ^{\left( p \right)} }
\right|_{\Gamma _0 } = 0;
\end{equation}
\begin{equation}
\label{eq4} \left[ {\dot {E}_\tau ^{\left( p \right)} }
\right]_\Gamma = \left[ {\dot {H}_\tau ^{\left( p \right)} }
\right]_\Gamma = 0.
\end{equation}
The point over functions shows that we use another definition of
eigenwaves and associated waves (not in the sense of Definition
\ref{def01}). We assume that
\begin{equation}
\label{eq5} \dot {E}^{\left( p \right)} \equiv \dot {H}^{\left( p
\right)} \equiv 0, \quad p < 0.
\end{equation}

We can prove the equivalence of both definitions for sufficiently
smooth functions $\Pi _p $ and $\Psi _p $; i.e. prove the
equalities (the proof in detail one can find in \cite{162})

\[
\dot {E}_3^{\left( p \right)} = \Pi _p , \quad \dot {H}_3^{\left(
p \right)} = \Psi _p , \quad p = 0,\;1,\;\dots ,\;m,
\]
and
\begin{equation}
\dot {E}_1^{\left( p \right)} = E_1^{\left( p \right)} ,\quad
\dots ,\quad \dot {H}_3^{\left( p \right)} = H_3^{\left( p
\right)}\quad {\mbox { for all}} \quad p \ge 0.
\label{formula16}
\end{equation}

{\bf{Remark.}}
{\textit{If the multiplicity of eigenvalue $\gamma $ is greater
than 1, we choose $\dot {E}_3^{\left( p \right)}$ and $\dot
{H}_3^{\left( p \right)}$ equal to functions $\Pi_p $ and $\Psi
_p$. If the choice is different then equations (\ref{formula16})
are generally not valid; however, the subspaces of eigenwaves and
associated waves corresponding to eigenvalue $\gamma $ will be the
same.}}

Let us strengthen that associated waves (\ref{eq1}) are defined
only from longitudinal components $\Pi _p $ and $\Psi _p $, which
allows us to study pencil $L\left( \gamma \right)$. This was
actually a purpose of new Definition \ref{def01}.
\subsection{Completeness of system of transversal components
of eigenwaves and associated waves}
\label{subsec:mylabel2}
Define
the transversal components $E_t^{\left( p \right)} = \left(
{E_1^{\left( p \right)} ,E_2^{\left( p \right)} } \right)^{\rm T}$,
$H_t^{\left( p \right)} = \left( {H_1^{\left( p \right)}
,H_2^{\left( p \right)} } \right)^{\rm T}$ of eigenwaves and associated
waves corresponding to eigenvalue $\gamma $; the longitudinal
components are $\Pi _p $ and $\Psi _p $. Introduce the
differential operators
\[
\nabla f = \frac{\partial f}{\partial x_1 }{\rm {\bf e}}_1 +
\frac{\partial f}{\partial x_2 }{\rm {\bf e}}_2 , \quad {\nabla
}'f = \frac{\partial f}{\partial x_2 }{\rm {\bf e}}_1 -
\frac{\partial f}{\partial x_1 }{\rm {\bf e}}_2 .
\]

We will prove the following formulas:
\begin{equation}
\label{eq16}
 - i\int\limits_\Omega {\left( {E_t^{\left( p \right)} \left( { - \nabla
'\bar {g}} \right) + H_t^{\left( p \right)} \left( {\nabla '\bar
{f}} \right)} \right)dx} = \int\limits_\Omega {\left( {\varepsilon
\Pi _p \bar {f} + \Psi _p \bar {g}} \right)dx},
\end{equation}
\begin{equation}
\label{eq17}
 - i\int\limits_\Omega {\left( {\varepsilon E_t^{\left( p \right)} \left(
{\nabla \bar {f}} \right) + H_t^{\left( p \right)} \left( {\nabla
\bar {g}}
\right)} \right)dx}
 = \gamma \int\limits_\Omega {\left( {\varepsilon \Pi _p \bar {f} + \Psi _p
\bar {g}} \right)dx} + \int\limits_\Omega {\left( {\varepsilon \Pi
_{p - 1}
\bar {f} + \Psi _{p - 1} \bar {g}} \right)dx} ; \\
\end{equation}
for
$p = 0,\;1,\;\dots ,\;m, \quad \forall f \in H_0^1
\left( \Omega \right), \quad g \in
\mathord{\buildrel{\lower3pt\hbox{$\scriptscriptstyle\frown$}}\over
{H}} ^1\left( \Omega \right)$.

Components $E_t^{\left( p \right)} $ and $H_t^{\left( p \right)} $
are defined by (\ref{eq1}). Using
induction with respect to $p$ it is easy to verify that
$E_t^{\left( p \right)} $ and $H_t^{\left( p \right)} $ satisfy
equations (\ref{eq2}). In particular,
\[
\frac{\partial E_2^{\left( p \right)} }{\partial x_1 } -
\frac{\partial E_1^{\left( p \right)} }{\partial x_2 } +
iH_3^{\left( p \right)} = \frac{\gamma }{\tilde {k}^2}\left(
{\frac{\partial E_2^{\left( {p - 1} \right)} }{\partial x_1 } -
\frac{\partial E_1^{\left( {p - 1} \right)} }{\partial x_2 }}
\right)
 + \frac{i}{\tilde {k}^2}\left( {\Delta \Psi _p + \tilde {k}^2\Psi _p }
\right) +
\]
\[
+ \frac{1}{\tilde {k}^2}\left( {\frac{\partial H_1^{\left(
{p - 1} \right)} }{\partial x_1 } + \frac{\partial H_2^{\left( {p
- 1} \right)} }{\partial x_2 }} \right)
 = \frac{i}{\tilde {k}^2}\left( {\Delta \Psi _p + \tilde {k}^2\Psi _p }
\right)
 + \frac{2\gamma }{\tilde {k}^2}\left( {\frac{\partial E_2^{\left( {p - 1}
\right)} }{\partial x_1 } - \frac{\partial E_1^{\left( {p - 1}
\right)} }{\partial x_2 }} \right) +
\]
\[
+ \frac{1}{\tilde {k}^2}\left(
{\frac{\partial E_2^{\left( {p - 2} \right)} }{\partial x_1 } -
\frac{\partial E_1^{\left(
{p - 2} \right)} }{\partial x_2 }} \right) =
\frac{i}{\tilde {k}^2}\left( {\Delta \Psi _p + \tilde {k}^2\Psi _p -
2\gamma \Psi _{p - 1} - \Psi _{p - 2} } \right) = 0;
\]
\[
\frac{\partial H_2^{\left( p \right)} }{\partial x_1 } -
\frac{\partial H_1^{\left( p \right)} }{\partial x_2 } -
i\varepsilon E_3^{\left( p \right)} = - \frac{i\varepsilon
}{\tilde {k}^2}\left( {\Delta \Pi _p + \tilde {k}^2\Pi _p }
\right) -
\]
\[
 - \frac{\varepsilon }{\tilde {k}^2}\left( {\frac{\partial E_1^{\left( {p -
1} \right)} }{\partial x_1 } + \frac{\partial E_2^{\left( {p - 1}
\right)} }{\partial x_2 }} \right) + \frac{\gamma }{\tilde
{k}^2}\left( {\frac{\partial H_2^{\left( {p - 1} \right)}
}{\partial x_1 } - \frac{\partial H_1^{\left( {p - 1} \right)}
}{\partial x_2 }} \right) =
\]
\[
 = - \frac{i\varepsilon }{\tilde {k}^2}\left( {\Delta \Pi _p + \tilde
{k}^2\Pi _p } \right) + \frac{2\gamma }{\tilde {k}^2}\left(
{\frac{\partial H_2^{\left( {p - 1} \right)} }{\partial x_1 } -
\frac{\partial H_1^{\left( {p - 1} \right)} }{\partial x_2 }}
\right) +
\]
\[
 + \frac{1}{\tilde {k}^2}\left( {\frac{\partial H_2^{\left( {p - 2} \right)}
}{\partial x_1 } - \frac{\partial H_1^{\left( {p - 2} \right)}
}{\partial x_2 }} \right) = - \frac{i\varepsilon }{\tilde
{k}^2}\left( {\Delta \Pi _p + \tilde {k}^2\Pi _p - 2\gamma \Pi _{p
- 1} - \Pi _{p - 2} } \right) = 0.
\]

Since $\Pi _p $ and $\Psi _p $ are solutions of Helmholtz
equations with a smooth right-hand side, these
functions are infinitely smooth in $\Omega _1 $ and $\Omega _2 $.
Thus $E_t^{\left( p \right)} $, $H_t^{\left( p \right)} $ are also
infinitely smooth functions  in $\Omega _1 $ and $\Omega _2 $. The
verification of equations (\ref{eq2}) is not complicated. Thus,
\begin{equation}
\label{eq18} \frac{\partial E_2^{\left( p \right)} }{\partial x_1
} - \frac{\partial E_1^{\left( p \right)} }{\partial x_2 } = -
i\Psi _p ;
\end{equation}
\begin{equation}
\label{eq19} \frac{\partial H_2^{\left( p \right)} }{\partial x_1
} - \frac{\partial H_1^{\left( p \right)} }{\partial x_2 } =
i\varepsilon \Pi _p ; \quad p = 0,\;1,\;\dots ,\;m.
\end{equation}
Since
\[
\left. {\Pi _p } \right|_{\Gamma _0 } = 0, \quad \left.
{\frac{\partial \Psi _p }{\partial n}} \right|_{\Gamma _0 } = 0,
\]
we can use (\ref{eq1}) to obtain
\begin{equation}
\label{eq20} \left. {E_\tau ^{\left( p \right)} } \right|_{\Gamma
_0 } = 0, \quad \left. {H_n^{\left( p \right)} } \right|_{\Gamma
_0 } = 0; \quad p = 0,\;1,\;...,\;m.
\end{equation}

Let us verify the transmission conditions
\begin{equation}
\label{eq21} \left[ {E_\tau ^{\left( p \right)} } \right]_\Gamma =
0;
\end{equation}
\begin{equation}
\label{eq22} \left[ {H_\tau ^{\left( p \right)} } \right]_\Gamma =
0; \quad p = 0,\;1,\;...,\;m.
\end{equation}
Formulas (\ref{eq21}) and (\ref{eq22}) for eigenfunctions ($p =
0)$ directly follows from conditions (8) in \cite{SS1}.
 Assume that
(\ref{eq21}) and (\ref{eq22}) are valid for functions with the
indices $p = 0,\;1,\;\dots ,\;q - 1$. We will prove that these
formulas hold for $p = q$. From equations (\ref{eq2}) we obtain
\begin{equation}
\label{eq23} \left. {\left[ \varepsilon \right]E_\tau ^{\left( p
\right)} } \right|_\Gamma = - i\left[ {\frac{\partial \Psi _p
}{\partial n}} \right]_\Gamma ;
\end{equation}
\begin{equation}
\label{eq24} \left. {\left[ \varepsilon \right]H_\tau ^{\left( p
\right)} } \right|_\Gamma = i\left[ {\varepsilon \frac{\partial
\Pi _p }{\partial n}} \right]_\Gamma , \quad p = 0,\;1,\;...,\;q -
1
\end{equation}
(these formulas also follow from (\ref{eq1})). Then
\[
\left[ {E_\tau ^{\left( q \right)} } \right]_\Gamma = i\gamma
\left[ {\frac{1}{\tilde {k}^2}} \right]\left. {\frac{\partial \Pi
_q }{\partial \tau }} \right|_\Gamma + \gamma \left[
{\frac{1}{\tilde {k}^2}} \right]\left. {E_\tau ^{\left( {q - 1}
\right)} } \right|_\Gamma
 - i\left[ {\frac{1}{\tilde {k}^2}\frac{\partial \Psi _q }{\partial n}}
\right]_\Gamma - \left[ {\frac{1}{\tilde {k}^2}} \right]\left.
{H_n^{\left( {q - 1} \right)} } \right|_\Gamma =
\]
\[
 = - i\left[ {\frac{1}{\tilde {k}^2}} \right]\left( {\left. {\frac{\partial
\Pi _{q - 1} }{\partial \tau }} \right|_\Gamma - \frac{2\gamma
}{\left[ \varepsilon \right]}\left[ {\frac{\partial \Psi _{q - 1}
}{\partial n}} \right]_\Gamma - \frac{1}{\left[ \varepsilon
\right]}\left[ {\frac{\partial \Psi _{q - 2} }{\partial n}}
\right]_\Gamma } \right)
 + \left[ {\frac{1}{\tilde {k}^2}} \right]\left( {\gamma \left. {E_\tau
^{\left( {q - 1} \right)} } \right|_\Gamma - \left. {H_n^{\left(
{q - 1} \right)} } \right|_\Gamma } \right) =
\]
\[
 = - i\left[ {\frac{1}{\tilde {k}^2}} \right]\left( {\left. {\frac{\partial
\Pi _{q - 1} }{\partial \tau }} \right|_\Gamma - \frac{2\gamma
}{\left[ \varepsilon \right]}\left[ {\frac{\partial \Psi _{q - 1}
}{\partial n}} \right]_\Gamma - \frac{1}{\left[ \varepsilon
\right]}\left[ {\frac{\partial \Psi _{q - 2} }{\partial n}}
\right]_\Gamma } \right) +
\]
\[
 + \left[ {\frac{1}{\tilde {k}^2}} \right]\left( {\left. { - \frac{2i\gamma
}{\left[ \varepsilon \right]}\left[ {\frac{\partial \Psi _{q - 1}
}{\partial n}} \right]_\Gamma + i\frac{\partial \Pi _{q - 1}
}{\partial \tau }} \right|_\Gamma - \frac{i}{\left[ \varepsilon
\right]}\left[ {\frac{\partial \Psi _{q - 2} }{\partial n}}
\right]_\Gamma } \right) = 0;
\]
\[
\left[ {H_\tau ^{\left( q \right)} } \right]_\Gamma = i\left[
{\frac{\varepsilon }{\tilde {k}^2}} \right]\left. {\frac{\partial
\Pi _q }{\partial n}} \right|_\Gamma + i\gamma \left[
{\frac{1}{\tilde {k}^2}} \right]\left. {\frac{\partial \Psi _q
}{\partial \tau }} \right|_\Gamma
 + \left[ {\frac{1}{\tilde {k}^2}} \right]\left. {\left( {\varepsilon
E_n^{\left( {q - 1} \right)} } \right)} \right|_\Gamma + \gamma
\left[ {\frac{1}{\tilde {k}^2}} \right]\left. {H_\tau ^{\left( {q
- 1} \right)} } \right|_\Gamma =
\]
\[
 = - i\left[ {\frac{1}{\tilde {k}^2}} \right]\left( {\left. {\frac{\partial
\Psi _{q - 1} }{\partial \tau }} \right|_\Gamma + \frac{2\gamma
}{\left[ \varepsilon \right]}\left[ {\varepsilon \frac{\partial
\Pi _{q - 1} }{\partial n}} \right]_\Gamma + \frac{1}{\left[
\varepsilon \right]}\left[ {\varepsilon \frac{\partial \Pi _{q -
2} }{\partial n}} \right]_\Gamma } \right)
 + \left[ {\frac{1}{\tilde {k}^2}} \right]\left( {\left. {\left(
{\varepsilon E_n^{\left( {q - 1} \right)} } \right)}
\right|_\Gamma + \gamma \left. {H_\tau ^{\left( {q - 1} \right)} }
\right|_\Gamma } \right) =
\]
\[
 = - i\left[ {\frac{1}{\tilde {k}^2}} \right]\left( {\left. {\frac{\partial
\Psi _{q - 1} }{\partial \tau }} \right|_\Gamma + \frac{2\gamma
}{\left[ \varepsilon \right]}\left[ {\varepsilon \frac{\partial
\Pi _{q - 1} }{\partial n}} \right]_\Gamma + \frac{1}{\left[
\varepsilon \right]}\left[ {\varepsilon \frac{\partial \Pi _{q -
2} }{\partial n}} \right]_\Gamma } \right) +
\]
\[
 + \left[ {\frac{1}{\tilde {k}^2}} \right]\left( {i\left. {\frac{\partial
\Psi _{q - 1} }{\partial \tau }} \right|_\Gamma + \frac{2i\gamma
}{\left[ \varepsilon \right]}\left[ {\varepsilon \frac{\partial
\Pi _{q - 1} }{\partial n}} \right]_\Gamma + i\left[ {\varepsilon
\frac{\partial \Pi _{q - 2} }{\partial n}} \right]_\Gamma }
\right) = 0.
\]

Now we can prove formulas (\ref{eq16}) and (\ref{eq17}). Applying
Green's formula and using (\ref{eq18})--(\ref{eq22}) we obtain
\[
\int\limits_\Omega {\left( {E_t^{\left( p \right)} \left( { -
\nabla '\bar {g}} \right) + H_t^{\left( p \right)} \left( {\nabla
'\bar {f}} \right)} \right)dx} =
\]
\[
 = \int\limits_\Omega {\left( {\left( {\frac{\partial E_1^{\left( p \right)}
}{\partial x_2 } - \frac{\partial E_2^{\left( p \right)}
}{\partial x_1 }} \right)\bar {g} + \left( {\frac{\partial
H_2^{\left( p \right)} }{\partial x_1 } - \frac{\partial
H_1^{\left( p \right)} }{\partial x_2 }} \right)\bar {f}}
\right)dx}
 = i\int\limits_\Omega {\left( {\varepsilon \Pi _p \bar {f} + \Psi _p \bar
{g}} \right)dx} ,
\]
which finally proves formula (\ref{eq16}).

Using (\ref{eq1}) we have
\[
 - i\int\limits_\Omega {\left( {\varepsilon E_t^{\left( p \right)} \nabla
\bar {f} + H_t^{\left( p \right)} \nabla \bar {g}} \right)dx} =
\]
\[
 = - i\gamma \int\limits_\Omega {\left( {E_t^{\left( p \right)} \left( { -
\nabla '\bar {g}} \right) + H_t^{\left( p \right)} \left( {\nabla
'\bar {f}} \right)} \right)dx}
 - i\int\limits_\Omega {\left( {E_t^{\left( {p - 1} \right)} \left( { -
\nabla '\bar {g}} \right) + H_t^{\left( {p - 1} \right)} \left(
{\nabla '\bar {f}} \right)} \right)dx + }
\]
\[
 + \int\limits_\Omega {\left( {\frac{\partial \Pi _p }{\partial x_2
}\frac{\partial \bar {g}}{\partial x_1 } - \frac{\partial \Pi _p
}{\partial x_1 }\frac{\partial \bar {g}}{\partial x_2 } +
\frac{\partial \Psi _p }{\partial x_1 }\frac{\partial \bar
{f}}{\partial x_2 } - \frac{\partial \Psi _p }{\partial x_2
}\frac{\partial \bar {f}}{\partial x_1 }} \right)dx} =
\]
\[
 = - i\gamma \int\limits_\Omega {\left( {E_t^{\left( p \right)} \left( { -
\nabla '\bar {g}} \right) + H_t^{\left( p \right)} \left( {\nabla
'\bar {f}} \right)} \right)dx}
 - i\int\limits_\Omega {\left( {E_t^{\left( {p - 1} \right)} \left( { -
\nabla '\bar {g}} \right) + H_t^{\left( {p - 1} \right)} \left(
{\nabla '\bar {f}} \right)} \right)dx} ,
\]
from which it follows that formulas (\ref{eq17}) also hold.

Let $L_2^2 \left( \Omega \right)$ be the Cartesian product of two
copies of space $L_2 \left( \Omega \right)$.

\textbf{Lemma 1}. \textit{
\label{lem5} For any element $u \in L_2^2 \left( \Omega \right)$
the decomposition $ u = \varepsilon \nabla f + {\nabla }'g $ holds
for certain functions $f \in H_0^1 \left( \Omega \right)$ and $g
\in\mathord{\buildrel{\lower3pt\hbox{$\scriptscriptstyle\frown$}}\over
{H}} ^1\left( \Omega \right)$. }

{\bf{Proof.}} Define a function $f \in H_0^1 \left( \Omega
\right)$ from the variational relation
\begin{equation}
\label{eq25} \int\limits_\Omega {u\nabla \bar {\varphi }dx} =
\int\limits_\Omega {\varepsilon \nabla f\nabla \bar {\varphi }dx}
, \quad \forall \varphi \in H_0^1 \left( \Omega \right).
\end{equation}
By the Riesz theorem \cite{109} there exists the unique element
$f$. Denote
$
v: = u - \varepsilon \nabla f; \quad v \in L_2^2 \left( \Omega
\right).
$
Thus for any $\varphi \in C_0^\infty \left( \Omega_j \right) \subset
H_0^1 \left( \Omega \right)$, $j = 1,2$, by the definition of
generalized derivatives
\[
\int\limits_\Omega {\left( {\dvr\,v} \right)\bar {\varphi }dx} = -
\int\limits_\Omega {v\left( {\nabla \bar {\varphi }} \right)dx} =
- \int\limits_\Omega {u\nabla \bar {\varphi }dx} +
\int\limits_\Omega {\varepsilon \nabla f\nabla \bar {\varphi }dx}
= 0,
\]
and hence, $\dvr\,v = 0$ in $\Omega _j $, $j = 1,2$, in terms of
distributions. Since $v \in L_2^2 \left( {\Omega _j } \right)$,
$\dvr\,v = 0$ in $\Omega _j $ then \cite{157} the trace of the normal
component of the vector exists on piecewise smooth boundary
$\partial \Omega _j $:
\[
\left. {v \cdot n} \right|_{\partial \Omega _j } \in H^{ - 1 /
2}\left( {\partial \Omega _j } \right), \quad j = 1,2.
\]
From formula (\ref{eq25}) we find that element $f \in H_0^1 \left(
\Omega \right)$ is a solution to the problem
\[
\left\{ {\begin{array}{l}
 \varepsilon \Delta f = \mbox{div}\,u\;\mbox{in}\;\Omega_j ,j = 1,2; \\
 \left. f \right|_{\Gamma _0 } = 0,\,\,\left[ {\varepsilon \frac{\partial
f}{\partial n} - u \cdot n} \right]_\Gamma = 0. \\
 \end{array}} \right.
\]

The second transmission condition is equivalent to the condition
$\left[ {v \cdot n} \right]_\Gamma = 0$. Then for any $\varphi \in
C_0^\infty \left( \Omega \right) \subset H_0^1 \left( \Omega
\right)$ we have
\[
\int\limits_\Omega {\left( {\mbox{div}\,v} \right)\bar {\varphi
}dx} = - \int\limits_\Omega {v\left( {\nabla \bar {\varphi }}
\right)dx} + \int\limits_\Gamma {\left[ {v \cdot n} \right]_\Gamma
\bar {\varphi }d\tau } = - \int\limits_\Omega {u\nabla \bar
{\varphi }dx} + \int\limits_\Omega {\varepsilon \nabla f\nabla
\bar {\varphi }dx} = 0,
\]
 i.e. $\mbox{div}\,v = 0$ in $\Omega $ as a distribution.

Define an element $g \in
\mathord{\buildrel{\lower3pt\hbox{$\scriptscriptstyle\frown$}}\over
{H}} ^1\left( \Omega \right)$ by the variational relation
\begin{equation}
\label{eq26} \int\limits_\Omega {v{\nabla }'\bar {h}dx} =
\int\limits_\Omega {\nabla g\nabla \bar {h}dx} , \quad \forall h
\in {\rm P},
\end{equation}
where
$
{\rm P}: = \{ {h:\left. h \right|_{\Omega _j } \in C^1\left(
{\bar {\Omega }_j } \right),\;\;j = 1,2,\;\;\left[ h
\right]_\Gamma = 0,\;\;\int\limits_\Omega {hdx} = 0} \}.
$
By the Riesz theorem there exists the unique element $g \in
\mathord{\buildrel{\lower3pt\hbox{$\scriptscriptstyle\frown$}}\over
{H}} ^1\left( \Omega \right)$ since the left-hand side of
(\ref{eq26}) is an antilinear continuous functional on
$\mathord{\buildrel{\lower3pt\hbox{$\scriptscriptstyle\frown$}}\over
{H}} ^1\left( \Omega \right)$, and set $P$ is dense in
$\mathord{\buildrel{\lower3pt\hbox{$\scriptscriptstyle\frown$}}\over
{H}} ^1\left( \Omega \right)$.

Every element $w \in Q$, where
$
Q: = \left\{ {w:w \in C^1\left( {\bar {\Omega }_j } \right),j =
1,2,\left[ w \right]_\Gamma = 0} \right\},
$
can be represented in the form $w = \nabla p + {\nabla }'h,$ where
$p \in \left\{ {p:p \in {\rm P},\left. p \right|_{\Gamma _0 } = 0}
\right\}$, $h \in {\rm P}$. Since functions $p$ and $h$ are smooth,
this relation can be proved in a standard manner using curve
integrals \cite{60}.

Taking into account the condition $\mbox{div }v = 0$, we see that
(\ref{eq26}) is equivalent to the variation relation
\begin{equation}
\label{eq27} \int\limits_\Omega {v \cdot \bar {w}dx} =
\int\limits_\Omega {{\nabla }'g \cdot \bar {w}dx} , \forall w \in
Q.
\end{equation}
However $Q$ is dense in $L_2^2 \left( \Omega \right)$, hence $v =
{\nabla }'g$, which proves the lemma.

\textbf{Lemma 2}. \textit{
\label{lem6} For any element $u \in L_2^2 \left( \Omega \right)$
the decomposition $ u = {\nabla }'f + \nabla g, $ holds for
certain $f \in H_0^1 \left( \Omega \right)$ and $g \in
\mathord{\buildrel{\lower3pt\hbox{$\scriptscriptstyle\frown$}}\over
{H}} ^1\left( \Omega \right).$ }

{\bf{Proof.}} The proof is similar to that of Lemma 1.
Element $f \in H_0^1 \left( \Omega \right)$ is defined by the
variational relation
\begin{equation}
\label{eq28} \int\limits_\Omega {u{\nabla }'\bar {\varphi }dx} =
\int\limits_\Omega {\nabla f\nabla \bar {\varphi }dx} , \quad
\forall \varphi \in H_0^1 \left( \Omega \right).
\end{equation}
Setting
$
v: = u - \nabla 'f, \quad v \in L_2^2 \left( \Omega \right),
$
we find that $\frac{\partial v_1 }{\partial x_2 } - \frac{\partial
v_2 }{\partial x_1 } = 0$ in $\Omega $ as a distribution.

Element $g \in H_0^1 \left( \Omega \right)$ is defined from the
relation
\[
\int\limits_\Omega {v \cdot \nabla \bar {p}dx} =
\int\limits_\Omega {\nabla g\nabla \bar {p}dx} , \quad \forall p
\in \left\{ {p:p \in {\rm P},\left. p \right|_{\Gamma _0 } = 0}
\right\},
\]
which is equivalent to
\[
\int\limits_\Omega {v \cdot \bar {w}dx} = \int\limits_\Omega
{\nabla g \cdot \bar {w}dx} , \quad \forall w \in Q,
\]
so that $v = \nabla g$.

Let $L_2^4 \left( \Omega \right)$ denote the Cartesian product of
four copies of space $L_2 \left( \Omega \right)$. Let $\left(
{E_{n,t}^{\left( p \right)} ,H_{n,t}^{\left( p \right)} }
\right)^{\rm T}$ be the transversal components of eigenwaves and
associated waves corresponding to eigenvalues $\gamma _n $, $p =
0,\;1,\;\dots ,\;m_n $. By $A$ we denote a set of indices
$n \in A$ assuming that different eigenvectors $\varphi _0^{\left(
n \right)} $ have different indices $n$ and the case $\gamma _n =
\gamma _m $ for $n \ne m$ is possible.

\textbf{Theorem 4}. \textit{
\label{th8} Let $\varepsilon _1 \ne \varepsilon _2 $. If the
system of eigenvectors and associated vectors $\left\{ {\varphi
_p^{\left( n \right)} } \right\}(p = 0,\;1,\;\dots ,\;m_n )$ of
pencil $L\left( \gamma \right)$ corresponding to eigenvalues
$\gamma_n $, $n \in A$, is double complete in $H\times H$, then
the system of vector-functions $\left\{ {\left( {E_{n,t}^{\left( p
\right)} ,H_{n,t}^{\left( p \right)} } \right)^{\rm T}} \right\}$,
$n \in A$, $p = 0,\;1,\;\dots ,\;m_n $, is complete in $L_2^4
\left( \Omega \right).$ }

{\bf{Proof.}} Using Lemmas 1 and 2
we
represent an arbitrary element $u \in L_2^4 \left( \Omega \right)$
in the form
\[
u = \left( {\begin{array}{l}
 \varepsilon \nabla f_2 - {\nabla }'g_1 \\
 {\nabla }'f_1 + \nabla g_2 \\
 \end{array}} \right)\,{\begin{array}{*{20}c}
 {f_j \in H_0^1 \left( \Omega \right),} \hfill \\
 {g_j \in
\mathord{\buildrel{\lower3pt\hbox{$\scriptscriptstyle\frown$}}\over
{H}}
^1\left( \Omega \right);} \hfill \\
\end{array} }\;{\begin{array}{*{20}c}
 \hfill \\
 {j = 1,2.} \hfill \\
\end{array} }
\]
It is sufficient to show that from the conditions
\begin{equation}
\label{eq29} \int\limits_\Omega {\left( {E_{n,t}^{\left( p
\right)} \left( {\varepsilon \nabla \bar {f}_2 - {\nabla }'\bar
{g}_1 } \right) + H_{n,t}^{\left( p \right)} \left( {{\nabla
}'\bar {f}_1 + \nabla \bar {g}_2 } \right)} \right)dx} = 0,\quad n \in A, \quad p = 0,\;1,\;\dots ,\;m_n ,
\end{equation}
we have $u = 0$.

By formulas (\ref{eq16}) and (\ref{eq17}) equations (\ref{eq29})
are reduced to the equations
\[
\int\limits_\Omega {\left( {\varepsilon \Pi _p^{\left( n \right)}
\bar {f}_1 + \Psi _p^{\left( n \right)} \bar {g}_1 } \right)dx} +
\gamma _n \int\limits_\Omega {\left( {\varepsilon \Pi _p^{\left( n
\right)} \bar {f}_2 + \Psi _p^{\left( n \right)} \bar {g}_2 }
\right)dx} +
\]
\begin{equation}
\label{eq30}
 + \int\limits_\Omega {\left( {\varepsilon \Pi _{p - 1}^{\left( n \right)}
\bar {f}_2 + \Psi _{p - 1} \bar {g}_2 } \right)dx} = 0, \quad n
\in A, \quad p = 0,\;1,\;\dots ,\;m_n ,
\end{equation}
or, in the operator form,
\begin{equation}
\label{eq31} \left( {K\varphi _p^{\left( n \right)} ,\tilde {f}_0
} \right) + \left( {\gamma _n K\varphi _p^{\left( n \right)} +
K\varphi _{p - 1}^{\left( n \right)} ,\tilde {f}_1 } \right) = 0,
\quad n \in A, \quad p = 0,\;1,\;\dots ,\;m_n ,
\end{equation}
where $\varphi _p^{\left( n \right)} = \left( {\Pi _p^{\left( n
\right)} ,\Psi _p^{\left( n \right)} } \right)^{\rm T}$, $\tilde {f}_0 =
\left( {f_1 ,g_1 } \right)^{\rm T}$, and $\tilde {f}_1 = \left( {f_2
,g_2 } \right)^{\rm T}$.

Taking into account (\ref{eq28w}),
one can show that (\ref{eq31}) is equivalent to the equations
\[
\left( {K\varphi _p^{\left( {n,v} \right)} ,\tilde {f}_v } \right)
= 0; \quad v = 0,\;1; \quad n \in A, \quad p = 0,1,...,m_n ,
\]
where $\varphi _p^{\left( {n,0} \right)} = \varphi _p^{\left( 0
\right)} $ and $\varphi _p^{\left( {n,1} \right)} = \gamma _n
\varphi _p^{\left( n \right)} + \varphi _{p - 1}^{\left( n
\right)} $. In view of  $K > 0$ and using the double completeness
of system $\left\{ {\varphi _p^{\left( n \right)} } \right\}$ in
$H\times H$ we obtain that
\[
\left( {\varphi _p^{\left( {n,v} \right)} ,K\tilde {f}_v } \right)
= 0, \quad n \in A, \quad p = 0,\;1,\;\dots ,\;m_n ,
\]
$K\tilde {f}_v = 0$, and $\tilde {f}_v = 0$, $v = 1,2$.

By this theorem the proof of completeness of the system of
transversal components of eigenwaves and associated waves in
$L_2^4 \left( \Omega \right)$ is reduced to the proof of double
completeness of the system of eigenvectors and associated vectors
of pencil $L\left( \gamma \right)$ in $H\times H$. This problem
was considered in Section \ref{section3}, where we established
sufficient conditions for the double completeness of the system of
eigenvectors and associated vectors of pencil $L\left( \gamma
\right)$ in $H\times H$. Thus under the conditions of Theorems 2
and 3
the system of transversal components
of eigenwaves and associated waves is complete in $L_2^4 \left(
\Omega \right)$.

In Theorem 4
we used the assumption $\varepsilon _1 \ne
\varepsilon _2 $. If $\varepsilon _1 = \varepsilon _2 $ then the
problem on normal waves is reduced to two scalar Dirichlet and
Neumann problems for the Helmholtz equation.
\subsection{Biorthogonal property for eigenwaves and associated waves}
\label{subsec:mylabel3}
Introduce the following notations. Let

\qquad $V_n^{\left( p \right)} : = \left( {E_{n,t}^{\left( p
\right)} ,H_{n,t}^{\left( p \right)} } \right)^{\rm T}$ and $W_n^{\left(
p \right)} : = \left( {H_{n,t}^{\left( p \right)} \times {\rm {\bf
e}}_3 ,{\rm {\bf e}}_3 \times E_{n,t}^{\left( p \right)} }
\right)^{\rm T}$

\noindent be the transversal components of an eigenwave ($p = 0)$
or associated wave ($p \ge 1)$ for the 'direct' wave and
'conjugate' wave, respectively (corresponding to eigenvalue
$\gamma _n )$. Brackets $\left\langle { \cdot , \cdot }
\right\rangle $ denote the inner product in $L_2^4 \left( \Omega
\right)$:
\[
\left\langle {V,W} \right\rangle = \int\limits_\Omega {V \cdot
\bar {W}dx} .
\]

Let us prove the following basic formula:
\begin{equation}
\label{eq32} \left( {\gamma _n - \gamma _m } \right)\left\langle
{V_n^{\left( p \right)} ,W_m^{\left( q \right)} } \right\rangle =
\left\langle {V_n^{\left( p \right)} ,W_m^{\left( {q - 1} \right)}
} \right\rangle - \left\langle {V_n^{\left( {p - 1} \right)}
,W_m^{\left( q \right)} } \right\rangle ,\quad p \ge 0, \,\, q \ge
0,
\end{equation}
where $V_n^{\left( p \right)} \equiv 0$ and  $W_m^{\left( q
\right)} \equiv 0$ for $p < 0$, $q < 0$.

Components $E_{n,t}^{\left( p \right)} $, $H_{n,t}^{\left( p
\right)} $ satisfy equations (\ref{eq2}). Expressing $\gamma _n
V_n^{\left( p \right)} $ and $\gamma _m W_m^{\left( q \right)} $
from these equations and using (\ref{eq16}) and (\ref{eq17}) we
obtain
\[
\int\limits_\Omega {\left( {\gamma _n V_n^{\left( p \right)} }
\right) \cdot \bar {W}_m^{\left( q \right)} dx} -
\int\limits_\Omega {V_n^{\left( p \right)} \cdot \left( {\gamma _m
\bar {W}_m^{\left( q \right)} } \right)dx} =
\]
\[
 = - i\int\limits_\Omega {\left( {H_{m,t}^{\left( q \right)} \left( { -
{\nabla }'\Pi _p^{\left( n \right)} } \right) + E_{m,t}^{\left( q
\right)} \left( {{\nabla }'\Psi _p^{\left( n \right)} } \right) +
E_{n,t}^{\left( p \right)} \left( { - {\nabla }'\Psi _q^{\left( m
\right)} } \right) + H_{n,t}^{\left( p \right)} \left( {{\nabla
}'\Pi _q^{\left( m \right)} } \right)} \right)dx}
\]
\[
 = - \int\limits_\Omega {\left( {\varepsilon \Pi _q^{\left( m \right)} \Pi
_p^{\left( n \right)} + \Psi _q^{\left( m \right)} \Psi _p^{\left(
n \right)} } \right)dx} + \int\limits_\Omega {\left( {\varepsilon
\Pi _p^{\left( n \right)} \Pi _q^{\left( m \right)} + \Psi
_p^{\left( n \right)} \Psi _q^{\left( m \right)} } \right)dx} +
\]
\[
 + \left\langle {V_n^{\left( p \right)} ,W_m^{\left( {q - 1} \right)} }
\right\rangle - \left\langle {V_n^{\left( {p - 1} \right)}
,W_m^{\left( q \right)} } \right\rangle = \left\langle
{V_n^{\left( p \right)} ,W_m^{\left( {q - 1} \right)} }
\right\rangle - \left\langle {V_n^{\left( {p - 1} \right)}
,W_m^{\left( q \right)} } \right\rangle .
\]
By formula (\ref{eq32}) we have that for $p \ge 0$ and $q \ge 0$
the following relations hold:
\begin{equation}
\label{formula34} \left\langle {V_n^{\left( p \right)}
,W_m^{\left( q \right)} } \right\rangle = 0\quad {\mbox{for}}
\quad \gamma _n \ne \gamma _m.
\end{equation}

Note that eigenvalues with different indices may coincide:
$
\gamma _n = \gamma _m \quad {\mbox {for}} \quad n \ne m.
$

%
\textbf{Theorem 5}. \textit{
\label{th9} Let $\varepsilon_1 \ne \varepsilon_2$ and system of
eigenvectors and associated vectors $\left\{ {\varphi _p^{\left( n
\right)} } \right\}$ $\left( {p = 0,\;1,\;\dots ,\;m_n } \right)$
of pencil $L\left( \gamma \right)$ corresponding to eigenvalues
$\gamma _n $, $n \in A$, be double complete in $H\times H$. Then
the system of vector-functions $\left\{ {\left( {E_{n,t}^{\left( p
\right)} ,H_{n,t}^{\left( p \right)} } \right)^{\rm T}} \right\}$,
$n \in A, p = 0,\;1,\;\dots ,\;m_n $, is complete and satisfies
the `minimality' property in $L_2^4 \left( \Omega \right)$ and
there exists a unique biorthogonal system  to the system of
vector-functions. }

{\bf{Proof.}} Completeness of system $\left\{ {V_n^{\left( p
\right)} } \right\}$ was proved in Theorem 4
The
'minimality' follows from the existence of a biorthogonal system
\cite{152}.

Renumber $\gamma _n $ so that eigenvalues with different indices
do not coincide with each other: $\tilde {\gamma }_k $, $k \in
\tilde {A}$. Let us form the systems of functions

\begin{eqnarray*}
\Lambda \left( {\tilde {\gamma }_k } \right): &=&
\bigcup\limits_{n,p:\gamma _n = \tilde {\gamma }_k ,0 \le p \le
m_n } {V_n^{\left( p \right)} } , {\rm dim}\,L\left( {\tilde
{\gamma }_k } \right) = r_k; Q\left( {\tilde {\gamma }_k }
\right): = \\
&=& \bigcup\limits_{n,p:\gamma _n = \tilde {\gamma }_k ,0 \le p
\le m_n } {W_n^{\left( p \right)} } , {\rm dim}\,Q\left( {\tilde
{\gamma }_k } \right) = r_k.
\end{eqnarray*}

Taking into account (\ref{eq34}), we note that in order to
construct a system biorthogonal to $\left\{ {V_n^{\left( p
\right)} } \right\}$, $n \in A$, $p = 0,\;1,\;\dots ,\;m_n $, it
is sufficient to form finite biorthogonal systems to $\Lambda
\left( {\tilde {\gamma }_k } \right)$ for a fixed $K$ using
$Q\left( {\tilde {\gamma }_k } \right)$ and then combine these
systems. We will form elements of a biorthogonal system to
$\Lambda \left( {\tilde {\gamma }_k } \right)$ as linear
combinations of elements of $Q\left( {\tilde {\gamma }_k }
\right)$.

Let $v_1 ,\;\ldots ,\;v_r \;$ and $w_1 ,\;\ldots ,\;w_r \;$ be
elements of systems $\Lambda \left( {\tilde {\gamma }_k } \right)$
and $Q\left( {\tilde {\gamma }_k } \right)$, respectively. Find
$
u_q = \sum\limits_{p = 1}^r {\bar {a}_{pq} w_p } , \quad q =
1,\;\ldots ,\;r,
$
from the conditions
\begin{equation}
\label{eq33} \left\langle {v_p ,u_q } \right\rangle = \delta _{pq}
, \quad p,q = 1,\;\ldots ,\;r,
\end{equation}
which are equivalent to the matrix equation $GA = I$, where $A: =
\left\{ {a_{pq} } \right\}$ and $G: = \left\{ {\left\langle {v_p
,w_q } \right\rangle } \right\}$ are of order $r\times r$. Since
system $\left\{ {V_n^{\left( p \right)} } \right\}$ is complete in
$L_2^4 \left( \Omega \right)$ the determinant of matrix $G$ is not
equal to zero. Thus there exists the unique matrix $A = G^{ - 1}$.


For homogeneous waveguides ($\varepsilon _1 = \varepsilon _2 )$
with the piecewise smooth boundary of domain $\Omega $ and in the
absence of associated waves the relations similar to
(\ref{formula34}) are well-known \cite{63}. However the problem is
not a vector one in this case.
\subsection{On basis property of system of eigenwaves
and associated waves} \label{subsec:mylabel4} In view of Theorem
5,
a natural question arises concerning the basis property
of system $\left\{ {V_n^{\left( p \right)} } \right\}$ in $L_2^4
\left( \Omega \right)$. Below we perform the analysis in terms of
the Schauder basis \cite{76}.

Let us prove the following lemmas.

\textbf{Lemma 3}. \textit{
\label{lem7} Let  $\left\{ {\varphi _i } \right\}$ be a complete
normal system in Hilbert space $H$ $\left( {\left\| {\varphi _i }
\right\| = 1} \right)$ and system $\left\{ {\psi _j } \right\}$
satisfies the conditions $ \left( {\varphi _i ,\psi _j } \right) =
N_j \delta _{ij} $ and $ 0 < C_1 \le \left\| {\psi _j } \right\|
\le C_2 . $ If there exists a subsequence $N_{j_k }$ of sequence
$N_j$ such that $N_{j_k } \to 0$ for $j_k \to \infty $, then
$\left\{ \varphi_i  \right\}$ is not a basis in $H$. }

{\bf{Proof.}} Since system $\left\{ {\varphi _i } \right\}$ is
complete in $H$, $N_j \ne 0$. Let $\left\{ {\varphi _i } \right\}$
be a basis in $H$. Let us form the system ${\psi }'_j = {\psi _j }
\mathord{\left/ {\vphantom {{\psi _j } {N_j }}} \right.
\kern-\nulldelimiterspace} {N_j }$, which is biorthogonal to
$\left\{ {\varphi _i } \right\}$: $\left( {\varphi _i ,{\psi }'_j
} \right) = \delta _{ij} $. Then $\left\{ {{\psi }'_j } \right\}$
is also a basis in $H$ according to \cite{76}, p. 371. If basis
$\left\{ {\varphi _i } \right\}$ is normal, then basis $\left\{
{{\psi }'_j } \right\}$ is almost normal, \cite{76}, p. 372. We
have ${\displaystyle \left\| {{\psi }'_{j_k } } \right\| =
{\left\| {\psi _{j_k } } \right\|}/{N_{j_k } } \to \infty ,}$\quad
i.e. $N_{j_k } \to 0$, for $j_k \to \infty $. Thus we have a
contradiction to the condition of normalization of $\left\{ {{\psi
}'_j } \right\}$.

\textbf{Lemma 4}. \textit{
\label{lem8} Let  system $\left\{ {\varphi _i } \right\}$ be a
basis in Hilbert space $H$. Then system $\left\{ {{\varphi }'_i }
\right\}$, ${\varphi }'_i = {\varphi _i } \mathord{\left/
{\vphantom {{\varphi _i } {\left\| {\varphi _i } \right\|}}}
\right. \kern-\nulldelimiterspace} {\left\| {\varphi _i }
\right\|}$ is also a basis in $H$. }

The proof in detail  can be found in \cite{SS1}.

Consider eigenvalue $\gamma $ of multiplicity 1 and compute inner
product $\left\langle {V^{\left( 0 \right)},W^{\left( 0 \right)}}
\right\rangle $ for $\gamma \ne 0$ using (\ref{eq1}) and
(\ref{eq16}):
\[
\gamma \left\langle {V,W} \right\rangle = 2\gamma
\int\limits_\Omega {\left( {E_1 H_2 - E_2 H_1 } \right)dx} = -
\int\limits_\Omega {\left( {\varepsilon E_t^2 + H_t^2 } \right)dx}
+
\]
\[
\begin{array}{l}
 \\
 + \int\limits_\Omega {\left( {E_1 \left( {\varepsilon E_1 + \gamma H_2 }
\right) + E_2 \left( {\varepsilon E_2 - \gamma H_1 } \right) + H_1
\left( {H_1 - \gamma E_2 } \right) + H_2 \left( {H_2 + \gamma E_1
} \right)}
\right)dx} = \\
 \end{array}
\]
\[
 = - \int\limits_\Omega {\left( {\varepsilon E_t^2 + H_t^2 } \right)dx} +
i\int\limits_\Omega {\left( {E_t \left( { - {\nabla }'\Psi }
\right) + H_t \left( {{\nabla }'\Pi } \right)} \right)dx} =
\]
\[
 = - \int\limits_\Omega {\left( {\varepsilon E_t^2 + H_t^2 } \right)dx} -
\int\limits_\Omega {\left( {\varepsilon \Pi ^2 + \Psi ^2}
\right)dx} .
\]
Here $V \equiv V^{\left( 0 \right)}$ and $W \equiv W^{\left( 0
\right)}$. Thus for the eigenwaves corresponding to $\gamma _n \ne
0$ of multiplicity 1 we have
\begin{equation}
\label{eq34}
\left\langle {V_n^{\left( 0 \right)} ,W_n^{\left( 0 \right)} }
\right\rangle
 = - \frac{1}{\gamma _n }\int\limits_\Omega {\left( {\varepsilon \left(
{E_{n,t}^{\left( 0 \right)} } \right)^2 + \left( {H_{n,t}^{\left(
0 \right)} } \right)^2} \right)dx} - \frac{1}{\gamma _n
}\int\limits_\Omega {\left( {\varepsilon \left( {\Pi _0^{\left( n
\right)} } \right)^2 + \left( {\Psi _0^{\left( n \right)} }
\right)^2} \right)dx} .
\end{equation}

Similarly, for the eigenwaves of multiplicity 1 corresponding to
$\gamma $ such that $\gamma \ne \bar {\gamma }$, we obtain
\[
 \int\limits_\Omega {\left( {\varepsilon \left| {E_t } \right|^2 + \left|
{H_t } \right|^2} \right)dx} = - \bar {\gamma }\int\limits_\Omega {\left( {E_1 \bar {H}_2 - E_2 \bar {H}_1
- H_1 \bar {E}_2 + H_2 \bar {E}_1 } \right)dx} +
\]
\[
 + \int\limits_\Omega {\left( {E_1 \left( {\varepsilon \bar {E}_1 + \bar
{\gamma }\bar {H}_2 } \right) + E_2 \left( {\varepsilon \bar {E}_2
- \bar {\gamma }\bar {H}_1 } \right) + H_1 \left( {\bar {H}_1 -
\bar {\gamma }\bar {E}_2 } \right) + H_2 \left( {\bar {H}_2 + \bar
{\gamma }\bar {E}_1 }
\right)} \right)dx}  =
\]
\[
 = - i\int\limits_\Omega {\left( {E_t \left( { - {\nabla }'\bar {\psi }}
\right) + H_t \left( {{\nabla }'\bar {\Pi }} \right)} \right)dx} -
\bar {\gamma }\overline {\left\langle {\bar {V},W} \right\rangle }
= \int\limits_\Omega {\left( {\varepsilon \left| \Pi \right|^2 +
\left| \Psi \right|^2} \right)dx} .
\]

It follows from \cite{SS1}, Theorem 2 that $\bar {V}$ is a transversal
component of the eigenwave corresponding to $\bar {\gamma }$.
Using (\ref{eq32}) we obtain that $\left\langle {\bar {V},W}
\right\rangle = 0$ for $\gamma \ne \bar {\gamma }$. Thus, for
eigenwaves of multiplicity 1 corresponding to $\gamma _n \ne \bar
{\gamma }_n $ we have
\begin{equation}
\label{eq35} \int\limits_\Omega {\left( {\varepsilon \left|
{E_{n,t}^{\left( 0 \right)} } \right|^2 + \left| {H_{n,t}^{\left(
0 \right)} } \right|^2} \right)dx} = \int\limits_\Omega {\left(
{\varepsilon \left| {\Pi _0^{\left( n \right)} } \right|^2 +
\left| {\psi _0^{\left( n \right)} } \right|^2} \right)dx} .
\end{equation}

Lemmas 3 and 4
and formulas (\ref{eq34})
and (\ref{eq35}) allow us to prove the following statement.

\textbf{Theorem 6}. \textit{
\label{th10} Let $\varepsilon _1 \ne \varepsilon _2 $, the
spectrum of pencil $L\left( \gamma \right)$ contain an infinite
set of isolated eigenvalues $\gamma _n $ of multiplicity 1, $n \in
\tilde {A}$, and $\gamma _n \to \infty $ for $n \to \infty $. Then
the system of vector-functions $\left\{ {\left( {E_{n,t}^{\left( p
\right)},\, H_{n,t}^{\left( p \right)} } \right)^{\rm T}}
\right\}$ corresponding to the system of eigenvectors and
associated vectors $\left\{ {\varphi _p^{\left( n \right)} }
\right\}\left( {p = 0,\;1,\;\ldots ,\;m_n } \right)$ of pencil
$L\left( \gamma \right)$ corresponding to eigenvalues $\gamma _n ,
n \in A$, $\tilde {A} \subset A$ is not a basis in $L_2^4 \left(
\Omega \right).$ }

{\bf{Proof.}} Let system $\left\{ {V_n^{\left( p \right)} }
\right\}$, $p = 0,\;1,\;\dots ,\;m_n $, $n \in A$, be a basis in
$L_2^4 \left( \Omega \right)$. Define a normalized system $\left\{
{\mathord{\buildrel{\lower3pt\hbox{$\scriptscriptstyle\frown$}}\over
{V}} _n^{\left( p \right)} } \right\}$,
$\mathord{\buildrel{\lower3pt\hbox{$\scriptscriptstyle\frown$}}\over
{V}} _n^{\left( p \right)} : = {V_n^{\left( p \right)} }
\mathord{\left/ {\vphantom {{V_n^{\left( p \right)} } {\left\|
{V_n^{\left( p \right)} } \right\|}}} \right.
\kern-\nulldelimiterspace} {\left\| {V_n^{\left( p \right)} }
\right\|}$. From Lemma 4
it follows that this system is
also a basis in $L_2^4 \left( \Omega \right)$.

Using linear combinations of the elements of system $\left\{
{W_n^{\left( p \right)} } \right\}$, $p = 0,\;1,\;\dots ,\;m_n $,
$n \in A$, we define, similarly to the proof of Theorem 5,
a unique biorthogonal system $\left\{ {\tilde
{W}_n^{\left( p \right)} } \right\}$ to basis $\left\{
{\mathord{\buildrel{\lower3pt\hbox{$\scriptscriptstyle\frown$}}\over
{V}} _n^{\left( p \right)} } \right\}$. Set
\[
\mathord{\buildrel{\lower3pt\hbox{$\scriptscriptstyle\frown$}}\over
{W}} _n^{\left( p \right)} = \left\{ {\begin{array}{l}
 \tilde {W}_n^{\left( p \right)} ,\quad n \notin \tilde {A}, \\
 {W_n^{\left( 0 \right)} }/{\left\| {W_n^{_{\left( 0 \right)} } }
\right\|},\;n \in \tilde {A}\;\left( {p = 0} \right). \\
 \end{array}} \right.
\]
For $n,m \in \tilde {A}$ we have
$
\left\langle
{\mathord{\buildrel{\lower3pt\hbox{$\scriptscriptstyle\frown$}}\over
{V}} _n^{\left( 0 \right)}
,\mathord{\buildrel{\lower3pt\hbox{$\scriptscriptstyle\frown$}}\over
{W}} _m^{\left( 0 \right)} } \right\rangle = \delta _{nm} N_n ,
\quad N_n : = {\left\langle {V_n^{\left( 0 \right)} ,W_n^{\left( 0
\right)} } \right\rangle } \mathord{\left/ {\vphantom
{{\left\langle {V_n^{\left( 0 \right)} ,W_n^{\left( 0 \right)} }
\right\rangle } {\left\| {V_n^{\left( 0 \right)} } \right\|^2}}}
\right. \kern-\nulldelimiterspace} {\left\| {V_n^{\left( 0
\right)} } \right\|^2},
$
because $\left\| {W_n^{\left( p \right)} } \right\| = \left\|
{V_n^{\left( p \right)} } \right\|$.

From (\ref{eq34}) and (\ref{eq35}) for sufficiently large $n$ and
$\gamma _n \ne \bar {\gamma }_n $ (see Theorems 1-4 in \cite{SS1})
we have
\[
 \left| {N_n } \right| \le \varepsilon _{\max } \left| {\gamma _n }
\right|^{ - 1}\frac{\int\limits_\Omega {\left( {\varepsilon \left|
{E_{n,t}^{\left( 0 \right)} } \right|^2 + \left| {H_{n,t}^{\left(
0 \right)} } \right|^2} \right)dx} + \int\limits_\Omega {\left(
{\varepsilon \left| {\Pi _0^{\left( n \right)} } \right|^2 +
\left| {\psi _0^{\left( n \right)} } \right|^2} \right)dx}
}{\int\limits_\Omega {\left( {\varepsilon \left| {E_{n,t}^{\left(
0 \right)} } \right|^2 + \left| {H_{n,t}^{\left( 0 \right)}
} \right|^2} \right)dx} }
 = \frac{2\varepsilon _{\max }}{ \left| {\gamma _n } \right|}
\]
and the latter tends to zero for $n \to \infty $, $n \in \tilde
{A}$.

Systems $\left\{
{\mathord{\buildrel{\lower3pt\hbox{$\scriptscriptstyle\frown$}}\over
{V}} _n^{\left( p \right)} } \right\}$ and $\left\{
{\mathord{\buildrel{\lower3pt\hbox{$\scriptscriptstyle\frown$}}\over
{W}} _n^{\left( p \right)} } \right\}$ satisfy conditions of Lemma
4;
consequently, $\left\{
{\mathord{\buildrel{\lower3pt\hbox{$\scriptscriptstyle\frown$}}\over
{V}} _n^{\left( p \right)} } \right\}$ in not the basis in $L_2^4
\left( \Omega \right)$. This contradiction proves the theorem.

Conditions of Theorem 6
are typical in rectangular and circular waveguides with partial
filling. Theorem 6
shows however that the basis property
of system $\left\{ {V_n^{\left( p \right)} } \right\}$ is not
valid for these problems.

Basis property is important when the problems of excitation of
waveguides are considered. The basis property of system $\left\{
{V_n^{\left( p \right)} } \right\}$ is applied e.g. in \cite{63}.

Note that  system  $\left\{ {V_n^{\left( p \right)} } \right\}$
does not form a basis when $\varepsilon _1 = \varepsilon _2 $.
However using decomposition into $E$- and $H$-waves one can show
that a `basis of subspaces' exists and expansions \cite{63} are
valid.
When $\varepsilon _1 \ne \varepsilon _2 $ the problem cannot be
reduced to any scalar problems for $E$- and $H$-waves.
\section{Conclusion}
%
%
We have formulated the definition of eigenwaves and associated
waves of a waveguide in terms of eigenvectors and associated
vectors of a pencil
and have shown that this definition is equivalent to a standard
one employing solutions of the Maxwell equations. We have
established double completeness of the system of eigenvectors and
associated vectors of the pencil
with a finite defect or without a defect. Using the obtained spectral
properties of the pencil we have proved the
completeness of the
system of transversal components of eigenwaves and associated
waves
%
and obtained biorthogonality relations.
We have established the `minimality' of the system of transversal
components of eigenwaves and associated waves and shown that in
the general case this system is not a Schauder basis.

The results obtained in this work are of fundamental character for the mathematical theory of wave propagation in guides
and must be used when excitation
of nonhomogeneously filled waveguides is considered.

\end{document}